# Report of the Working Group on Strategic Exoplanet Initiatives with HST and JWST

Seth Redfield, Wesleyan University (Chair)
Natasha Batalha, NASA Ames
Björn Benneke, University of Montréal
Beth Biller, University of Edinburgh (STUC Chair)
Nestor Espinoza, Space Telescope Science Institute
Kevin France, University of Colorado
Quinn Konopacky, University of California, San Diego
Laura Kreidberg, Max Planck Institute for Astronomy
Emily Rauscher, University of Michigan
David Sing, Johns Hopkins University

Neill Reid, Space Telescope Science Institute (ex officio)
Elena Sabbi, Space Telescope Science Institute (ex officio)

# 1. Executive Summary

The discovery and characterization of exoplanets is a field that has grown tremendously over the last decades, fueled by the confluence of forces including advances in telescopes and instruments across a wide wavelength range, the interconnectedness of stars and planets drawing previously siloed subfields together, and the prospect of answering deeply profound scientific questions associated with the cosmic context of life in the universe. The Astro2020 Decadal Survey identified exoplanet science as a top priority and exoplanet scientific questions are among the drivers of future facilities. *JWST* and *HST* have provided transformative discoveries in the field of exoplanets, and based on the high demand from community GO proposals, remain at the cutting edge of exoplanet research.

This Working Group (WG) was charged with soliciting community feedback and evaluating the strategic planning for exoplanet science with *JWST* and *HST* given the high quality of exoplanet observations, the significantly lengthened mission lifetime for *JWST*, and the pronounced expansion of the field over the last decade. We were charged with identifying key science themes, providing recommendations on issues associated with optimal timing and scale of resources, as well as providing a recommended DDT concept achievable with 500 hours of *JWST* time. We received feedback from the community in the form of surveys and white papers. The ideas for future observations were diverse, broad, compelling, and abundant. They addressed essentially all of the key science questions in our field identified by the Astro2020 Decadal Survey.

We identified three key science themes that should be emphasized in future *JWST* and *HST* GO and Archival efforts. The first is to understand the prevalence and diversity of atmospheres on rocky-M dwarf worlds. This theme forms the basis of the motivation for our DDT Concept recommendation to do a thermal emission survey of rocky worlds and identify if and which of the rocky M-dwarf exoplanets have atmospheres. The second key science theme is to understand population-level trends of exoplanet atmospheres from sub-Neptunes to gas giants. This theme forms the motivation around being strategic and scaffolding the inevitable $10^4$ Hour JWST Exoplanet Survey that will arise from future GO programs. The third key science theme is to understand exoplanets in the context of their stellar environments. This theme highlights the holistic approach that is necessary to understand exoplanets and address pathways to habitability. It argues for a multiwavelength approach and identifies the synergy of *JWST* and *HST* in answering key science questions.

We recommend a DDT concept to survey the atmospheres of rocky-M dwarf exoplanets. M dwarfs provide a unique opportunity to characterize rocky and potentially temperate atmospheres with *JWST*. These targets have been the focus of much study and motivate the design of future missions and programs. It is critical to quickly survey a wide sample of such targets to ascertain if they indeed host significant atmospheres, i.e., define the cosmic shoreline, and to identify high priority targets for future follow-up. The identification of these high value targets is the first step toward determining if M star habitable zone planets could actually be habitable. We recommend a comprehensive survey of thermal emission from rocky worlds over a range of temperatures. It is important for this effort to occur early in the mission lifetime. A conclusive answer based on a large survey will avoid a protracted effort where the TAC may be hesitant to allocate the time required to characterize the coolest targets if the hottest targets do not show signs of atmospheres. Also, this program will quickly identify the highest priority targets so that the community has ample mission lifetime to propose significant follow-up observations.

In the context of strategic planning of exoplanet observations, it is useful to estimate the expected exoplanet observational commitment over *JWST's* lifetime. Given the current usage associated with exoplanets, extended over 20 cycles, it is anticipated that *JWST* will dedicate ~30,000 hours to exoplanet observations. Based on the population studies associated with exoplanet detection and atmospheric characterization in the early cycles and in the community white papers, a significant fraction of these efforts will be capable of addressing fundamental science questions across populations (in mass, temperature, stellar host, etc). We recommend efforts to support GO-driven programs that will contribute to this unprecedented data product of *JWST*. Of the ~30,000 hours of anticipated *JWST* full-mission time dedicated to exoplanets, we expect that 1/3 of it could, and perhaps inevitably would, form a comprehensive, high S/N, panchromatic, $10^4$ hour atmospheric survey of planets. Such an observational sample would be a legacy archive that would address a broad range of science questions across various populations of planets. It would also bridge the direct imaging and transit communities and involve a multitude of techniques to detect and characterize exoplanets. With some institutional scaffolding, in the form of research community support (e.g., inclusive working groups, meeting support) or proposal mechanisms (e.g., checkbox initiatives or time subsidies) we feel this sample could be built from GO-driven programs.

With respect to issues associated with optimal timing, the WG makes a couple recommendations and has identified two areas for further study. We recommend a proprietary period across multi-epoch observations. Issues of proprietary period significantly impact our community and we need a mechanism that is sensitive to the fact that many of these observations are multi-epoch in nature. We also recommend scheduling priority for programs that may be impacted by programmatic timeliness. Given the multi-cycle nature of many exoplanet studies and the critical and unique contribution of *HST*, such observations should be prioritized immediately. We identify two study items. The first is to explore the feasibility of simultaneous *JWST* and *HST* (UV) observations. The second is to explore the possibility of funding in advance of the start of observing programs to support analysis and model preparation, and personnel acquisition necessary to carry out the proposed work.

With respect to issues associated with scale of resource, the WG makes several recommendations. We recommend STScI provide dedicated and robust funding support for the recommended DDT program. We also recommend STScI provide inclusive, program management support to encourage and scaffold GO-driven programs that will contribute to an eventual $10^4$ Hour *JWST* Exoplanet Survey. We recommend support of *HST*, ground-based, and X-ray observations required for a holistic understanding of exoplanets. In particular, the UV capabilities of *HST* are critical to contextualizing the atmospheric characterization by *JWST*. Finally, we recommend providing the most advanced data products as possible, a necessary resource for projects with dynamical scheduling and/or hierarchical implementation. The multi-epoch nature of many exoplanet studies requires this functionality in order to make the most efficient use of *JWST* and *HST*.

Based on the community feedback and work to-date by researchers using *JWST* and *HST*, it is abundantly clear that *JWST* and *HST* will provide transformative exoplanet observations. They will continue to produce fundamental exoplanet discoveries while also producing a legacy database that will be utilized by researchers for generations to come. These recommendations are meant to kickstart community-driven GO efforts and support those efforts to produce a final database whose ultimate impact is greater than the sum of its parts. These recommendations also represent concrete steps in the pursuit of pathways to habitable worlds.

# 2. Background and Charge

## 2.1 The Charge

STScI convened the WG on Strategic Exoplanet Initiatives with *HST* and *JWST* in April 2023 and gave the following charges:

0. Solicit input from the community on key science areas that should be prioritized for *HST* or *JWST* observations;
1. Identify science themes that should be prioritized for future *HST* and *JWST* General Observer programs and/or Archival analyses, including potential *HST* multi-cycle programs;
2. Provide advice on the optimal timing for substantive follow-up observations and suggest mechanisms for enabling those observations;
3. Comment on the appropriate scale of resources likely required to support those programs
4. Develop a specific concept for a large-scale (~500 hours) Director's Discretionary exoplanet program to start implementation by *JWST* Cycle 3.

## 2.2 Timeliness of Charge

This is the right time for strategic planning around exoplanet science and an investment of observing time now will pay large dividends in terms of community-driven follow-up programs in later cycles. The demonstrated precision capabilities of *JWST*, a much longer anticipated mission lifetime, and the fast pace of exoplanet discovery have made it clear that *JWST* and *HST* will provide transformative observations that will significantly impact the field.

*JWST* is an amazing facility for exoplanet observations. Its broad, infrared, spectroscopic capabilities are key for probing multiple molecular atmospheric features, unlocking chemistry of abundant and life-associated molecules (e.g., those containing carbon and oxygen), as well as new chemistry (e.g., sulfur). *JWST's* thermal and pointing stability leads to ultra-precise transit spectrophotometry and sensitive direct imaging observations. *JWST's* large aperture provides high signal-to-noise in a short period of time, e.g., in a single exoplanet transit.

*HST* hosts the only operating UV spectrographs capable of characterizing stellar emission and planetary properties in this critical wavelength regime. A comparable facility is at least 15–20 years from operation. Coordinated measurements of the host star UV emission will be critical in interpreting *JWST* exoplanets observations with the ultimate goal of understanding the physical processes that drive the past evolution and current state of the observed planets.

Exoplanet science questions are a high priority in Astronomy. The Astro2020 Decadal Survey identified as top priorities the Science Theme: Worlds and Suns in Context and Priority Area: Pathways to Habitable Worlds. Table E.1 in the Decadal Survey highlights 17 key science questions regarding exoplanets and essentially all are addressable with *JWST* and *HST* observations. The questions in the category E-Q3 (How do habitable environments arise and evolve within the context of their planetary systems?) can only be addressed with a survey of rocky worlds. We recommend a DDT Concept program that would answer the fundamental question of which terrestrial worlds orbiting low mass stars have atmospheres. This will enable community follow-up on the highest value targets and provide vital information to future telescope designs and project implementation. The questions in the category E-Q2 (What are the properties of individual planets, and which processes lead to planetary diversity?) are answerable with a comprehensive exoplanet population survey. These questions, and many more, will be addressed with the accumulation of *JWST* and *HST* observations in the coming years. We offer recommendations that can facilitate and organize the activities of the community to create a transformative and lasting population survey to answer these fundamental questions. Finally, we reiterate the Astro2020 Decadal Survey's advice on how to maximize exoplanet science, which included acquiring a large sample size across a diverse stellar sample, utilizing direct imaging and UV capabilities, and measuring multiple spectroscopic bands and species. *JWST* and *HST* provide the capabilities to answer these high priority science questions.

The landscape of exoplanet science has been fundamentally transformed over the developmental lifetime of *JWST*. In 2005, after initial construction had begun, a Science Assessment Team reported on the science requirements that would drive further design of the instruments. Needless to say, the number and diversity of exoplanets increased dramatically in the intervening years leading up to *JWST's* launch. Figure 1 shows the sample of known exoplanets at the time of *JWST's* 2005 Science Assessment, while Figure 2 shows the current sample of exoplanets. The axes of these figures are associated with properties of the exoplanet, the planetary equilibrium temperature and the planetary mass. The size and color of the symbols are associated with properties of the exoplanet's host star, the size is related to the J-band magnitude (an important determinant of the observability of the exoplanet) and the color is associated with the stellar spectral type. The number of exoplanets on this figure grew from many dozens of planets to many thousands. The range of planets also was greatly expanded, with the emergence of super-Earths and sub-Neptunes, and the regular detection of terrestrial planets. With the new targets has come a host of new science questions, key among them, as prioritized by the Astro2020 Decadal Survey, the diversity of planetary properties and the habitability of terrestrial planets.

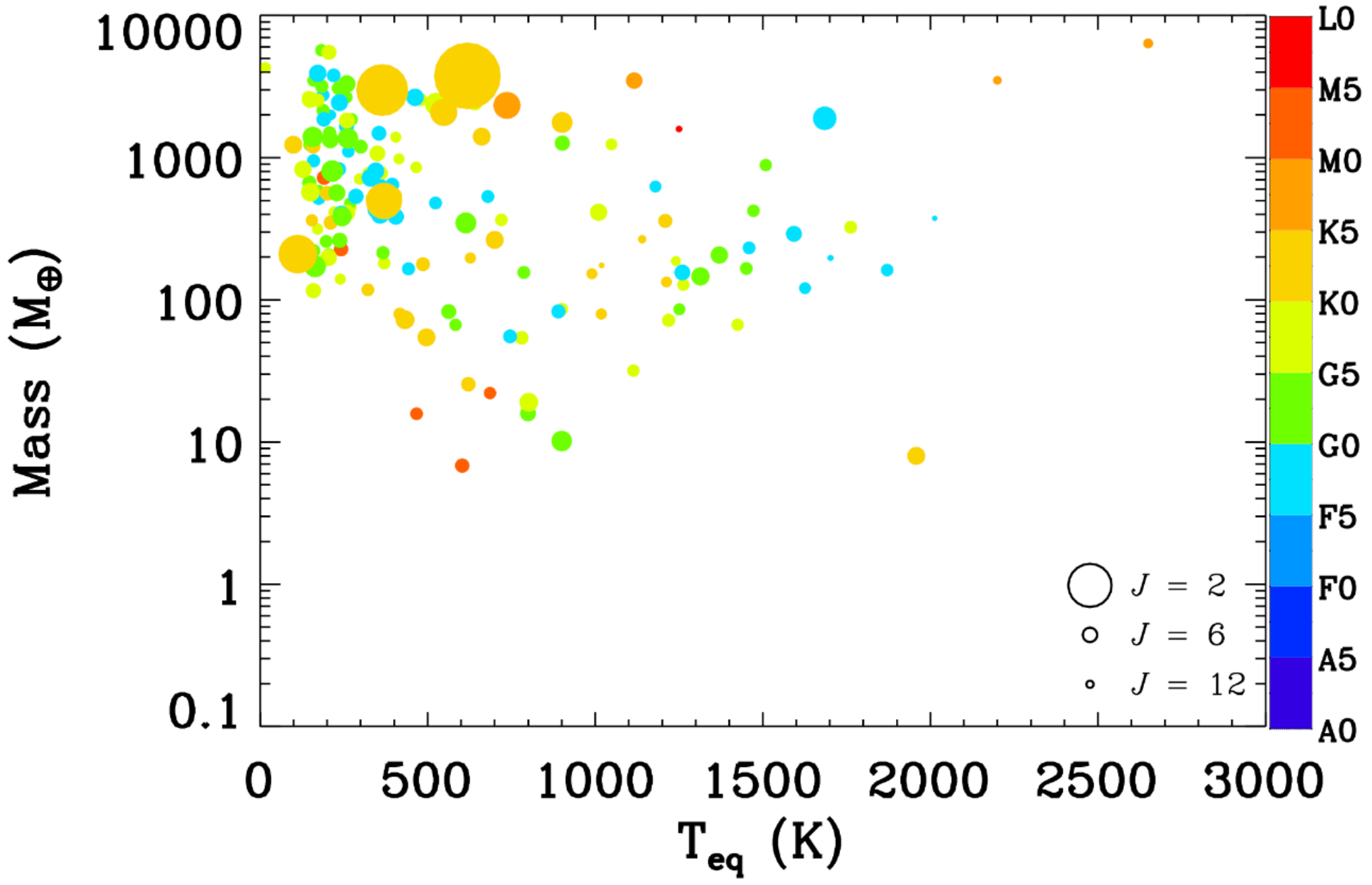

**Figure 1:** Exoplanet mass and equilibrium temperature of all known exoplanets in 2005 during *JWST's* Science Assessment. The symbol size is associated with the host star J-band magnitude and color with the spectral type of the host star.

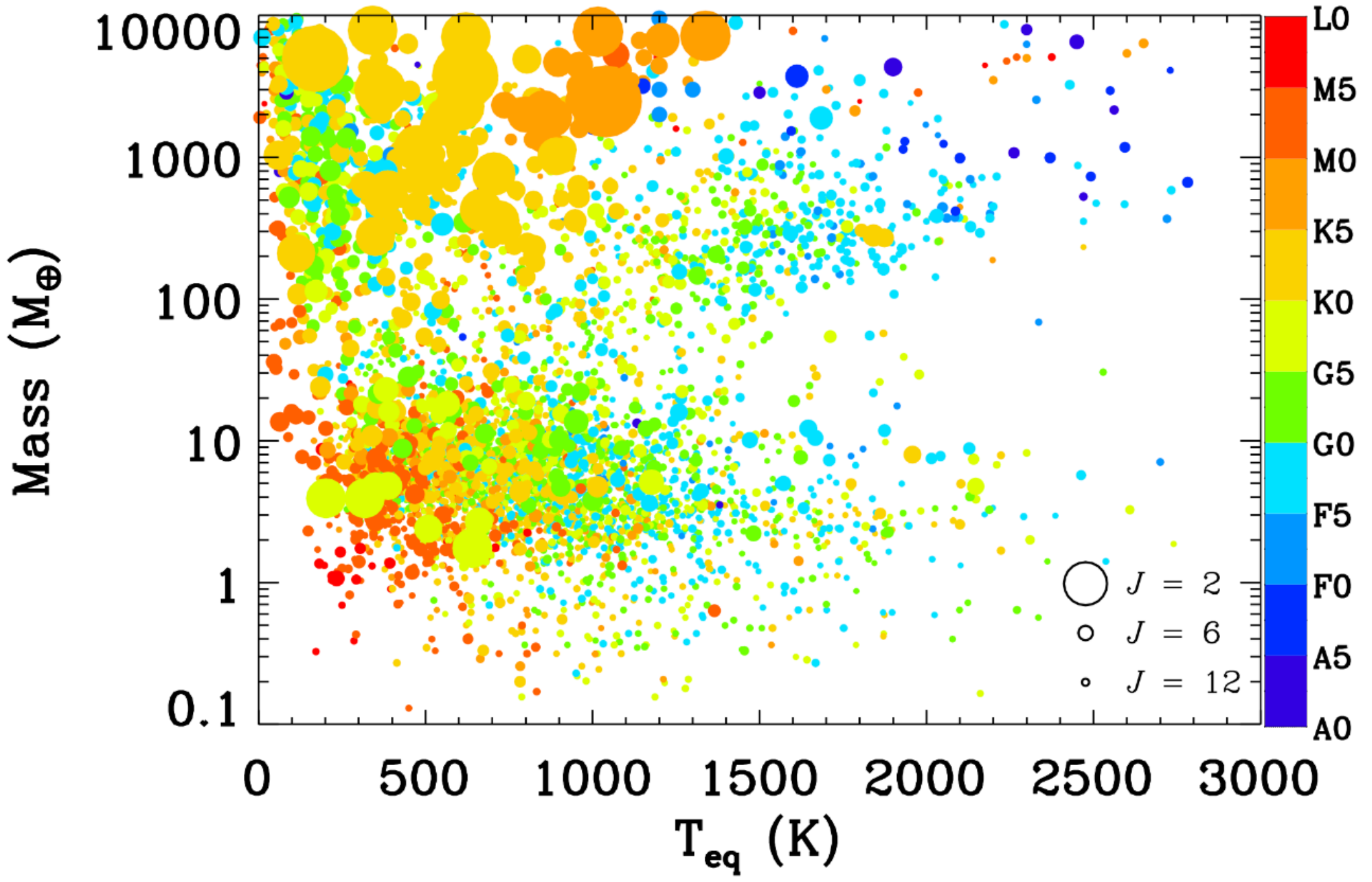

**Figure 2:** Exoplanet mass and equilibrium temperature of all known exoplanets in 2024 during the WG's deliberation. The symbol size is associated with the host star J-band magnitude and color with the spectral type of the host star.

## 2.3 Anticipating the $10^4$ hr *JWST* Exoplanet Survey by Cycle 20

Given the extended mission lifetime for *JWST* and the demonstrated time commitment to exoplanets and exoplanet formation in the initial cycles, *JWST* will have compiled a substantial archive of exoplanet observations by Cycle 20. For the purposes of strategic planning, it is useful to extrapolate the scale of the exoplanet commitment by *JWST*:

5000 hours per cycle * ⅓ allocated to exoplanets * 20 cycles * ⅓ tied to atmospheric characterization = $10^4$ hours.

This assumes 5000 hours allocated for GO efforts in each cycle, roughly a third of all time going to exoplanets and exoplanet formation (see Figure 3), that *JWST* will have 20 observing cycles, and that at least a third of the exoplanets and exoplanet formation time will be associated with programs to characterize the exoplanet atmosphere. This leads to $3 \times 10^4$ hours dedicated to exoplanets, driven by guest observers, and at least $10^4$ hours dedicated to atmospheric characterization of various populations of exoplanets across a range of possible physical properties. While it is impossible to predict the focus of these observations, it is likely that a large fraction (here we assume ⅓) will be of some value to answering questions of the diversity of exoplanet atmospheres and will include a large number of high S/N observations of the most easily observed systems, as well as some observations of systems that are extremely difficult and require pushing the unique capabilities of *JWST* to their limit. This sample will include observations across a large range in planet types and characteristics, obtained from several methods (e.g., transmission or emission spectroscopy in transit and direct imaging).

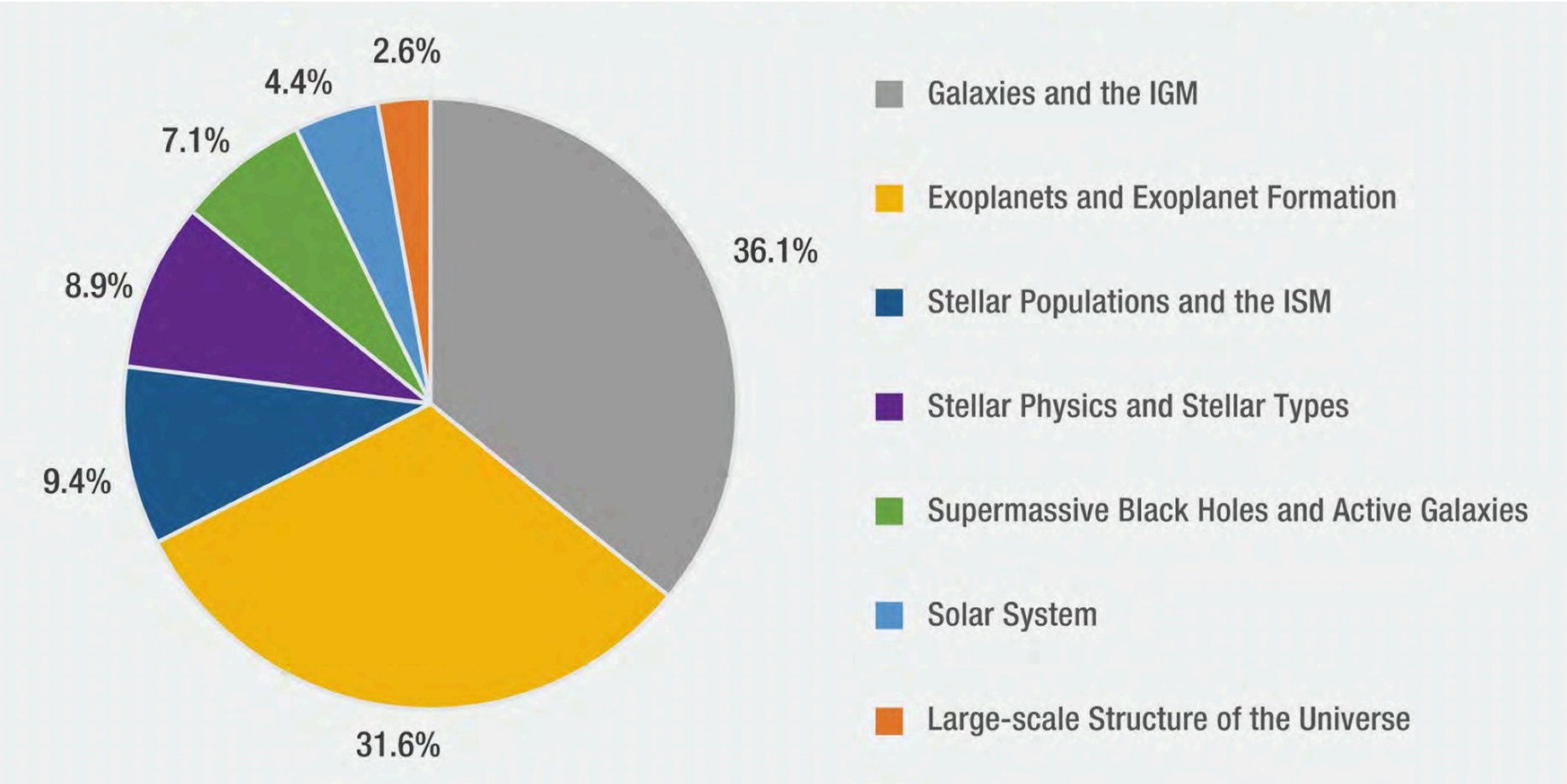

**Figure 3:** The fraction of time allocated by science category for *JWST* Cycle 3. Roughly one-third of all time is allocated to exoplanets and exoplanet formation. This figure was presented to the JSTUC as part of a presentation by the STScI Science Policies Group on Cycle 3 Preparations.

The scale of the anticipated *JWST* exoplanet sample coupled with the scale and synergy of the white papers submitted by the community feedback (which is likely to be replicated in the scale and synergy of GO proposals) presents an incredible strategic planning opportunity for our field. In this document, we provide recommendations in all charges to scaffold and support GO programs in future cycles to optimize this forthcoming $10^4$ hour *JWST* Exoplanet Survey which will include observations across a large range in planet types and characteristics, obtained from several methods (e.g., transmission or emission spectroscopy). In particular, the DDT Concept is an opportunity to kickstart scientific investigations that will inspire GO proposals in near-term cycles. By focusing on a specific part of the exoplanet parameter space envisioned in the $10^4$ hr survey, it will identify which atmospheres to follow-up on with deep dive spectroscopy and ensure that future cycle *JWST* exoplanet observations will answer key science questions in our field.

# 3. Community Engagement Process

An integral part of the WG charges involved obtaining community input. By design, it is this community input that defined the recommendations described in this report. Because of its importance, the WG took special care to engage the community via different means in as open a process as possible. This began what the WG refers to as the Community Engagement Process (CEP). The process for gathering community input was drafted and refined early in the WG discussions (April 2023), in order to initiate the community input gathering soon thereafter.

## 3.1 Process Definition

Over the course of two months (April–May 2023), the WG defined the ways in which the CEP would gather, compile and evaluate community input in light of the charges, and decided on four main avenues of participation from the community:

1. **White papers (WPs).** A 3-page document submitted anonymously (1 cover page, 1 page for text and 1 page for figures) providing input to any of the WG charges (key science themes, advice on optimal timing, appropriate scale of resources, specific concept for a large-scale DDT program). The WPs were read anonymously and authorship was revealed after the recommendations were in place. Authors were informed that the white papers would be made public when the report was finalized. The submission deadline was September 8, 2023. An overleaf template was provided, as well as a white paper with a fictitious topic for reference. The list of white paper submissions titles and authors are shown in Appendix A of this report.

2. **Survey.** A 3-page Google Form; page 1 covering demographic and prior *HST*/*JWST* use information, page 2 covering word-limited responses focusing on each of the WG charges and page 3 covering space for any additional feedback. Deadline for responses was September 8, 2023. The list of questions from the survey are shown in Appendix B.

3. **Town Halls.** Three, 1-hour, open virtual meetings were used to communicate the WG charges and CEP, and receive community input. The three took place on July 12, 19 and 31, 2023, set at different times to accommodate world-wide timezones. The last of those was specifically aimed at Early Career researchers (defined as student, postdoctoral researchers and research scientists within 10 years of their terminal degree). Meetings were run via BlueJeans, with Slido being used for Q&A.

4. **E-mail.** A special e-mail (wg-exoplanets@stsci.edu) was established for the WG to receive community input, both on questions about the CEP or general questions about the WG charges.

The WG created a public webpage (https://sites.google.com/view/exoplanet-strategy-wg/home) with all the information described above, including submission links for both the White Papers and the Survey that went live on June 23, 2023. This included a Frequently Asked Questions (FAQ) page which was updated with feedback received from the community throughout the process. In addition, WG members shared this information via various Slack channels (including the ones from the Early Release Science transiting exoplanet and high-contrast imaging teams) and e-mail lists, including the usage of the stscigeneric@stsci.edu e-mail list, from June through September 2023.

An important point of discussion in the WG was centered on how exactly the community input in the forms described above would be compiled in both recommendations to the STScI Director on the WG charges and to the community itself (via, e.g., this report). Throughout the process, the WG decided to follow two guiding principles:

1. **The WG is not a Time Allocation Committee (TAC)**; the WG would not attempt to rank survey or white paper responses for “winning ideas”. Rather, the WG would use those inputs to compile and aggregate community ideas.

2. **The WG will evaluate ideas**; the WG, entrusted with assessing and compiling ideas by the charge, would discuss and evaluate them with the goal of elevating those that best answered the charge (and not evaluate them on, e.g., just popularity of ideas).

## 3.2 Community Input Compilation and Evaluation

Using Google Analytics to measure WG webpage interaction with the community, the webpage had about 1,000 unique users throughout the CEP, with over 200 users repeatedly returning to the webpage. The three WG Town Halls had over 100 attendees each. E-mail reminders were sent a few days to a week prior to the Town Halls via the stscigeneric@stsci.edu e-mail list proving particularly effective at raising the number of interactions with the WG webpage. The questions asked during the Town Halls not addressed in the FAQ or in the WG charter were added as new FAQs on the webpage. We received a total of 5 e-mails to the special e-mail, all focused on questions about the process, some of which were also added to the FAQ.

A total of 42 white papers (WPs) and 75 survey answers were received. The full list of WPs, including authors, is shown in Appendix A of this report. More than 230 unique researchers were among the co-authors in the submitted WPs.

The demographic information obtained through the 75 survey responses that were received are presented in Figure 4. The majority of respondents seem to be white (70.1%), male (68.8%) coming from either US (50.6%) or European-based institutions (31.2%). In terms of career stage, tenured and non-tenured faculty and/or research scientists form the majority of the responses (65%) followed by responses from postdoctoral researchers (20.8%) and undergraduate and graduate students (9.1%). In terms of primary affiliation, the majority of the responses come from R1 or equivalent institutions and government or research institutions (74.1%). Finally, in terms of involvement on HST or JWST projects, there seems to be a more or less equal split between those that have only collaborated on projects involving those observatories (53.2%) and those that have *led* projects on them (46.8%).

**Table 1:** Percentage of surveys and WPs that addressed specific charges

| | % of surveys addressed | % of WPs addressed |
|---|---|---|
| Key Science Themes | 91% | 76% |
| Optimal Timing | 63% | 24% |
| Scale of Resources | 68% | 22% |
| DDT Concepts | 52% | 63% |

**Table 2**: Categories and percentage of DDT Concepts presented in the surveys and WPs

| | |
|---|---|
| Sub-Neptune to Giant broad atmospheric characterization surveys | 31% |
| Terrestrial/Sub-Neptune atmospheric characterization | 29% |
| Planet Searches | 22% |
| Stellar Characterization | 10% |
| Other | 6% |

The next step in the CEP was to collate and evaluate the input from the community from both the survey and the WPs. Following the guiding principles described in Section 3.1, the WG Chair defined leads tasked with "clustering" ideas present in all the forms of input described above. The leads then identified distinct ideas, mainly stemming from the survey and white papers, collating similar ideas drawn from different submissions. This clustering was performed during September 2023. High-level discussions were held both to ensure no idea or topic was missed, as well as to organize a detailed discussion of the input to each charge.

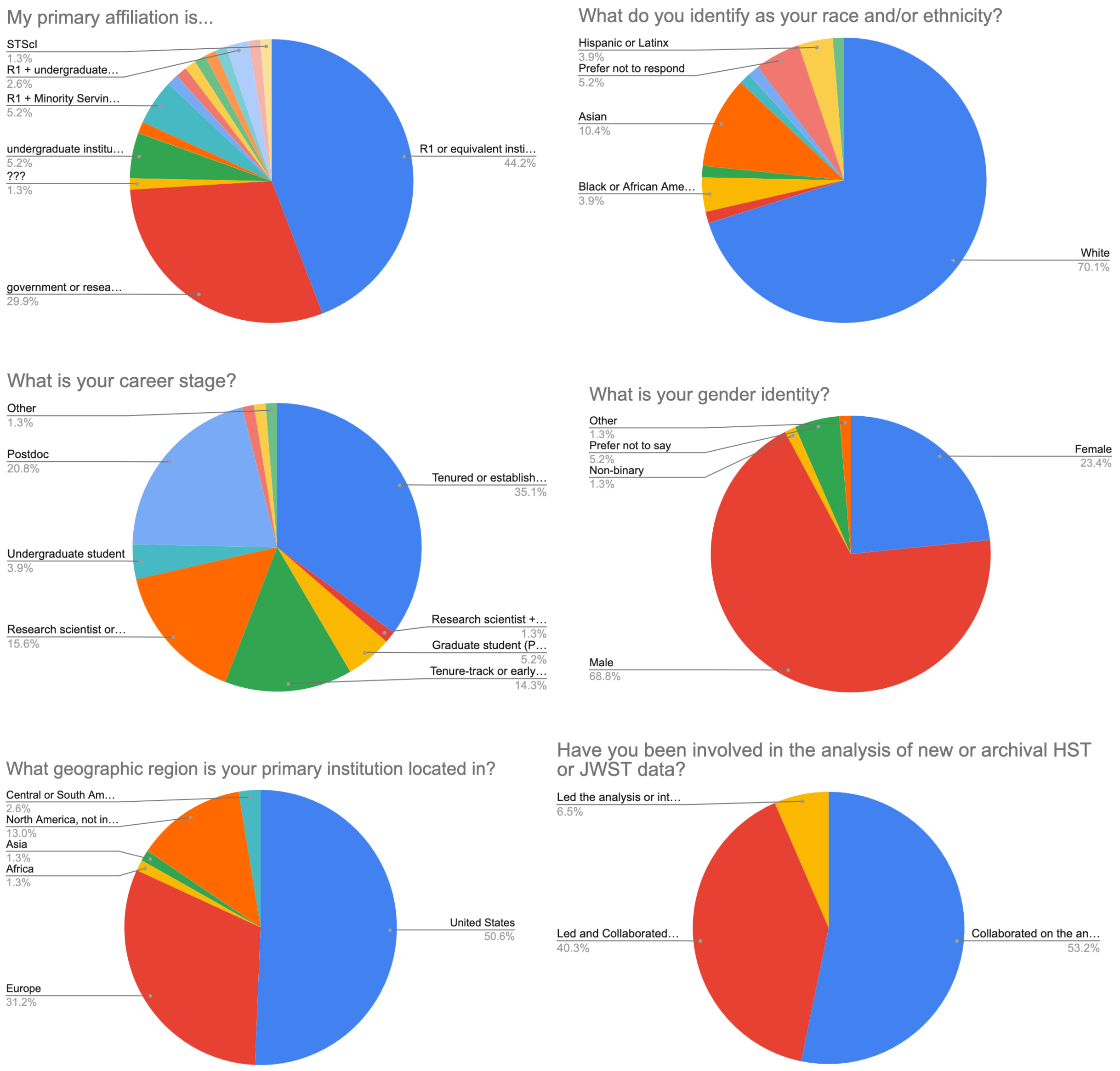

**Figure 4:** Demographic information from the 75 survey responses submitted by the community.

The decision to have both WPs and survey responses as possible inputs to the WG was based on both the targeted and relatively straightforward nature of surveys, as well as the lower time commitment related to answering survey questions. The WG believed that the survey responses would have a more homogeneous mix of topics being addressed, with the WPs likely addressing mostly key science themes and DDT Concepts. In Table 1 we show the percentage of responses that address the different charges for both the survey and the WPs, showing that this was indeed the case. Table 2 indicates a categorization of DDT Concepts and percentage of survey and WP submissions in each category.

The detailed discussion of those ideas was performed in an in-person meeting at STScI which was held on October 4–5, 2023. The majority of the meeting was devoted to Key Science Themes and DDT Concepts, as Optimal Timing and Scale of Resources had very specific and comprehensive inputs from the community. While the community input revealed a broad level of interest from the scientific community in terms of Key Science Themes and DDT Concepts on different exoplanet populations, a key aspect revealed by the clustering exercise was the division between “giant”, “sub-Neptune” and “rocky” exoplanet science, as well as separation of science themes by techniques (e.g., transits, high-contrast imaging, etc.). The details of this clustering are discussed further in Section 4 on Key Science Themes.

The DDT Concepts discussion focused on all ideas given as inputs to the WG via the survey and the white papers. This discussion led to identifying 6 main concepts prompted by the community:

1. A spectroscopic giant exoplanet survey to perform atmospheric observations of both transiting and direct imaging exoplanets in one or a few modes (large sample).

2. A spectroscopic giant exoplanet survey to perform atmospheric observations of both transiting and direct imaging exoplanets in all modes (small sample).

3. A photometric 15-micron rocky exoplanet survey focused on warm-to-temperate, rocky worlds orbiting M-dwarfs.

4. A sub-Neptune spectroscopic atmospheric survey.

5. A Direct Imaging search of planets around pre and main sequence stars.

6. A Direct Imaging search of planets around White Dwarfs.

The WG then used anonymous ranked choice voting to decide which concepts would ***not*** be kept in the discussion. In terms of first choice for a DDT concept, 60% of the votes went to Concept 3 (rocky M-dwarf survey) and 40% went to Concept 1 (large sample giant exoplanet survey). These two concepts were listed in every member’s top two choices.

The WG then decided to explore Concepts 3 and 1 in detail. To this end, the WG was divided in two, with each group tasked to develop and assess each concept more fully and present to the full WG an evaluation of each concept with respect to 8 key questions described in the solicitation to the community of concepts for the DDT program: (1) inability of the idea to be achieved via regular GO, (2) legacy/archival impact, (3) connection with the next generation exoplanet observatories, (4) maximizing returns of having *JWST* and *HST* operating contemporaneously, (5) urgency to do this early in *JWST’s* life, (6) risk/feasibility of the idea, (7) timeliness in terms of targets and key properties of the idea and (8) strength of community demand. Two internal documents were created with this discussion; after presentations about each idea were given to the entire WG, it was decided those documents would be reviewed in detail by all members, and a decision on our final recommendation would be made at a later time.

An initial presentation of the above preliminary discussion, along with a debrief of the material presented in the present report was delivered to the STScI Director and leadership team on November 15, 2023. At this time, it was made clear that the WG still had not decided between the rocky M-dwarf exoplanet (Concept 3) or the giant exoplanet survey idea (Concept 1) as our final recommendation. In a virtual meeting on November 29, 2023, the WG, after hearing once again the presentations of each of the ideas, held an anonymous vote to decide which would be the official WG recommendation. The rocky M-dwarf exoplanet survey idea (Concept 3) was selected with 90% of the votes. More details of the DDT Concept recommendation is given in Section 7.

# 4. Recommendations for Key Science Themes

The recommendations submitted by the community for key science themes spanned a diverse array of topics, many of which have also been highlighted as high priority by the Astro2020 Decadal Survey. Submissions with similar themes often recommended different observing strategies for the DDT program. Nevertheless, it was clear which major science themes the community thought should be given priority for the DDT program, and which should be prioritized for the next decades of *HST* and *JWST* programs. We emphasize that the working group was tasked with developing a specific concept for a large-scale (~500 hours) DDT program as well as identifying science themes that should be prioritized for future General Observer and/or Archival analyses. Figure 5 shows a synopsis of these different DDT concept ideas, survey submissions, and white papers submissions organized into aggregated science themes, highlighting the interconnectedness of various topics.

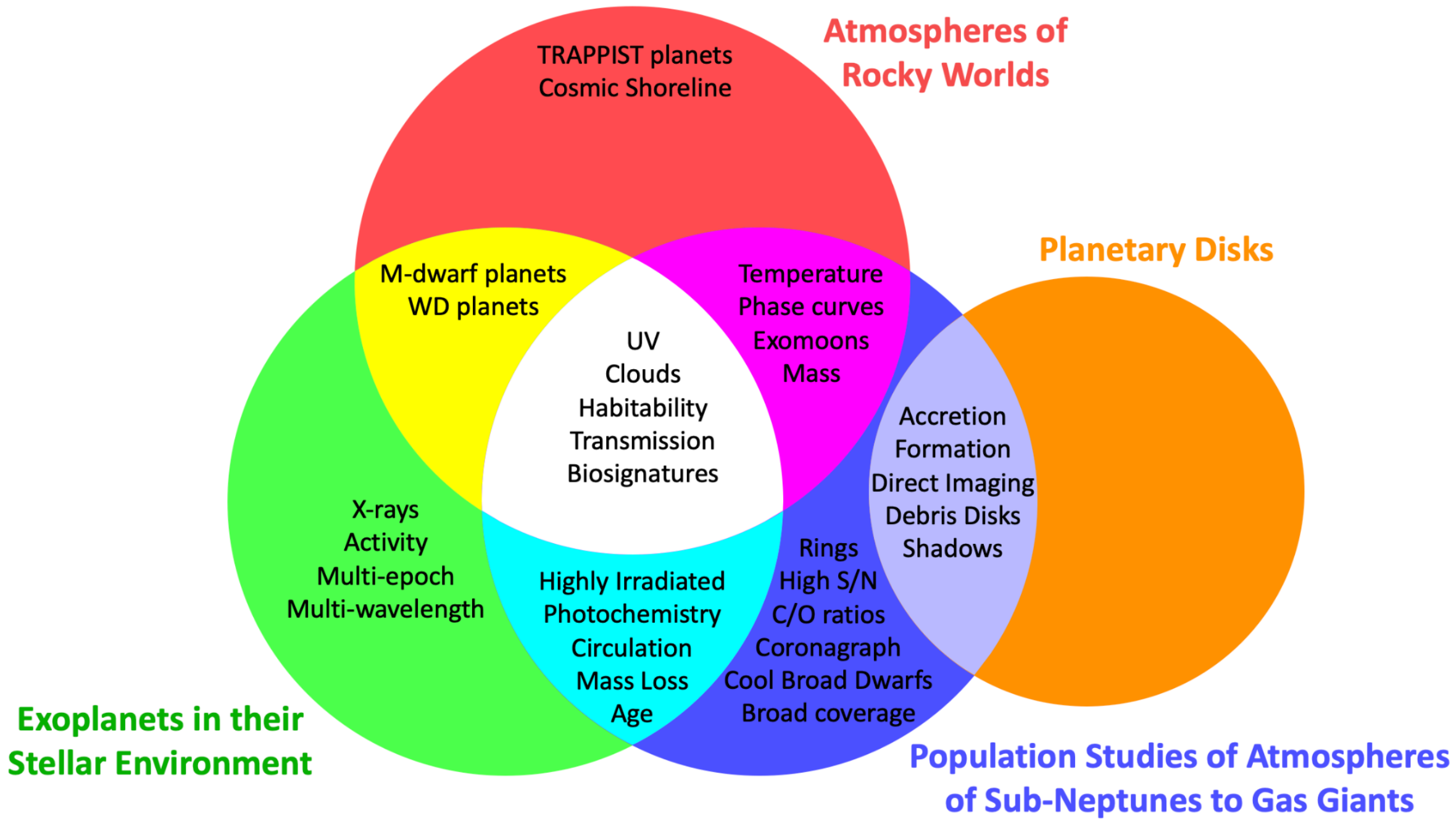

**Figure 5:** Summary of topics submitted by the community and their interrelationships.

In what follows we provide an overview of the key recommendations put forth for science themes by the community. In Section 7 we convey how the recommended DDT concepts connect back to these recommended science themes.

**Science Theme 1: Understand the prevalence and diversity of atmospheres on rocky-M dwarf worlds**

One of the common high-priority science themes that arose in the community feedback was the nature of small, rocky planet atmospheres around M dwarfs. The motivation for doing so centered around three fundamental results. First, the *Kepler* mission, ground-based surveys, and the *Transiting Exoplanet Survey Satellite (TESS)* have shown that rocky planets are relatively common in the Milky Way (e.g., Dattilo et al. 2023). Second, approximately 75% of all main-sequence stars in the Milky Way are M dwarfs, while only ~20% are FGK, making rocky planets around M-dwarfs the most numerous rocky planet in the galaxy (e.g., Dressing et al. 2013, Kopparapu et al. 2013). And third, *JWST* is uniquely suited to characterize the atmospheres of these planets in both transit and eclipse geometry, even when compared with future observatories such as the *Habitable Worlds Observatory (HWO)*, because of the optimization of *HWO* for characterization of planets around solar type stars.

There were three major topics that the community felt could be meaningfully addressed with *JWST* and *HST* observations:

**1.1 Understanding the prevalence of detectable atmospheres on rocky-M-dwarf planets**

Many observations of rocky-M-dwarf planets are already underway with *JWST* and *HST*. At the time of writing (January 2024), thermal emission has been measured for TRAPPIST-1b and c, and GJ 376b (Greene et al. 2023, Zieba et al. 2023, Zhang et al. 2024). These early results are consistent with no atmospheric circulation from TRAPPIST-1b or GJ 376b. TRAPPIST-1c is slightly cooler than a zero albedo bare rock, hinting at a thin atmosphere or a higher albedo surface. The first rocky planet transmission spectra have been measured as well (Lustig-Yaeger et al. 2022, Moran et al. 2023, May et al. 2023, Lim et al. 2023, Kirk et al. 2024). Yet, to date, no spectral features have been identified in transmission. These initial findings show that substantially more observational investment is needed to identify whether rocky M-dwarf planets can maintain atmospheres. Specifically, there was keen interest in understanding the population of rocky-M-dwarf planets, beyond detecting individual planet atmospheres.

Of central focus in the survey responses was the concept of the cosmic shoreline. The cosmic shoreline is an empirical relationship inspired by the Solar System that separates bodies more likely to have an atmosphere from those less likely to have an atmosphere based on their predicted atmospheric escape velocities and cumulative planet XUV insolation (Zahnle & Catling 2017). The community recognized that meaningfully testing the concept of a “cosmic shoreline” would require observations of many rocky-M-dwarf planets to sample parameter space across the properties of the planet (e.g., mass, radius, insolation, temperature) and star (e.g., M type, stellar activity).

This question touches on a major focus of the Astro2020 Decadal Survey to understand what processes, such as atmospheric escape, influence the habitability of environments. Therefore, ultimately the community was particularly motivated to continue the search for detectable atmospheres on rocky-M-dwarf planets.

### 1.2 Probing the diversity of atmospheres on rocky-M-dwarf planets

There was also community interest in conducting panchromatic observations of rocky-M-dwarf planets to understand the diversity of their atmospheres. This, of course, would require more in-depth and time-consuming observations, when compared to what may be needed to simply detect the atmosphere. Of particular focus was the desire to understand the relative abundances of common molecular species, such as $H_2O$, $CO_2$, $CH_4$. The presence of molecules such as $CO_2$, could be tracers for observable characteristics of habitable planets such as surface liquid water (e.g., Turbet et al. 2016, Charnay et al. 2015). Understanding relative abundances of $CO_2$ and $CH_4$ could shed light on the redox state of the atmosphere, which would allow us to place other rocky worlds in the context of our own terrestrial planets (e.g., Catling & Kasting 2007). This science theme transcended just potentially habitable planets and touched on investigating terrestrial worlds more broadly, including planets such as lava worlds and super-Earths. Overall, investing time to probe the diversity of atmospheres would enable the community to identify key observable characteristics of rocky M-dwarf planets and would push toward the characterization of habitable planets, a major focus of the Astro2020 Decadal Survey.

### 1.3 Searching for biosignatures and other signs of habitability on rocky-M-dwarf planets

There were several white papers and survey responses that emphasized the importance of tackling habitable exoplanets, especially given the focus of exoplanet habitability in the Astro2020 Decadal Survey. It was generally acknowledged that this type of survey would require dedicated spectroscopic observations of just a few key terrestrial systems (e.g., TRAPPIST-1, high priority temperate super-Earths/water worlds), which have been the major focus of planetary habitability thus far (e.g., Meadows 2018). Spectroscopic surveys would search for both signatures of habitability and biological activity. Therefore, this time would be well-invested given the unique opportunity for transformational science. Additionally, strategic investment in key habitable systems may lead to an understanding of what habitable exoplanets should be the focus of future studies throughout *JWST's* lifetime and for future missions.

Overall these three science topics associated with this key science theme are deeply linked to the Astro2020 Decadal Survey Priority Area: Pathways to Habitable Worlds. Tackling all three topics surrounding rocky-M-dwarf planets would give the community an understanding of the prevalence of nearby potential habitable planets, the characteristics of potential habitable planetary systems, the processes that influence the habitability of environments, and ultimately the key observable characteristics of habitable planets.

Sample of White Papers associated with this Science Theme: 15, 16, 21, 25, 26, 28, 29, 36, 38

**Science Theme 2: Understand population-level trends of exoplanet atmospheres from sub-Neptune to gas giants**

Following the first observations of exoplanets with H/He-dominated atmospheres using *JWST*, it is clear that the facility is ushering in a new era of unprecedented atmospheric characterization. This population of planets is vast and varied, ranging in size from sub-Neptunes to super-Jupiters, in temperature from a few hundred to several thousands of Kelvin, and in age from newborn to ancient. The key to unlocking the secrets of exoplanet evolution lies in the atmosphere, with subtle fingerprints pointing to formation pathways and dynamical history. At the same time, the detailed and complex physics that shapes atmospheres must be understood in order to make progress in explaining the observable properties of gaseous planets.

The importance of these areas of investigation was clearly expressed by the community in the white papers and surveys returned to this committee. Over 30% of the feedback for DDT Concepts focused on atmospheric characterization of H/He-dominant atmospheres, see Table 2. Here we coalesce that feedback into two broad categories of science goals: understanding the atmospheres in their own right, and then using the properties of the atmosphere to shed light on other facets of planetary evolution. There were two major topics that the community felt could be meaningfully addressed with *JWST* and *HST* observations:

**2.1 Understanding the physical and chemical processes that sculpt planetary atmospheres**

*JWST's* access to the full suite of near-to-mid infrared wavelengths at multiple spectral resolutions enables a new, detailed look at exoplanet atmospheric physics. Access to the fundamental bands of molecular species such as CO, $CH_4$, and $NH_3$ offer the ability to assess the relative strengths of these features comprehensively, probing properties such as atmospheric mixing and disequilibrium chemistry. In particular, mixing is generally parameterized in atmosphere models by the vertical mixing parameter $K_{zz}$ which remains poorly constrained but fundamentally shapes multiple atmospheric properties. One such property is the presence and composition of clouds, which are simultaneously ubiquitous and challenging to model. Multiwavelength data from *JWST* offers a unique look at cloud species, with the possibility of detecting spectral features from the clouds at longer wavelengths, yielding previously inaccessible compositional information (Miles et al. 2023). Another critical process sculpted by properties like mixing is upper atmosphere photochemistry. Photochemical processes leading to the development of hazes, which have already been shown to strongly impact observations. Of particular interest is photochemistry that results in the development of prebiotic molecules, particularly amongst the smaller sub-Neptune population (Moran et al. 2020). This process is ideally investigated by a combination of long wavelength observations from *JWST* and near-simultaneous data in the UV from *HST*. The sub-Neptune population shares potential atmospheric evolution overlap with the rocky worlds, including atmospheric loss, but are easier to observe than terrestrial planets.

The sensitivity of *JWST* also enables exploration of the three-dimensional nature of exoplanetary atmospheres. Full phase curve observations and eclipse mapping of gas giant planets provide detailed information about the physics of global circulation, enabling atmospheric mapping across the extent of the planet. Such complex models of planetary atmospheres rely on descriptions of clouds, composition, and temperature and wind structures, all of which are critical to understanding the population of gas giants as a whole (Showman et al. 2020). Monitoring directly imaged planetary mass companions for spectroscopic variability can also resolve longitudinal structure as a function of atmospheric depth and time (Bowler et al. 2020).

Indeed, an exciting prospect of this new era of atmospheric characterization is the ability to compare planets of varying type and evolutionary stages and thereby map the dependency of the atmospheric properties on fundamental parameters. Direct imaging surveys with *JWST* will image the first sub-Jupiter mass planets and add to the sample for follow-up observations, again with *JWST*. By comparing atmospheres as a function of mass, age, and level of irradiation, the field will move from the realm of individual discovery into that of statistical understanding of atmospheres. This requires comparing observables that span the technique boundary: observations of transiting/eclipsing planets should be compared to directly imaged systems in the same wavelength range. Furthermore, isolated very low mass objects that push into the planetary mass regime can yield excellent information about similar atmospheres without the complexities of removal or modeling of starlight (e.g., Hood et al. 2024).

### 2.2 Using atmospheric tracers to understand planetary formation, migration, accretion, and evolution

A major goal of exoplanet atmosphere science is to connect atmospheric chemical abundance patterns to the formation history of the planet. Abundances of Solar System planets have been attributed to factors such as the location of formation, migration history, and accretion of specific types of small bodies (e.g., Mousis et al. 2009). With precision abundances enabled by new measurements with *JWST*, there is now the potential to perform similar studies in exoplanetary systems. For instance, one of the canonical tracers of formation process and/or location is the carbon-to-oxygen ratio (e.g., Oberg et al. 2011, Piso et al. 2016, Madhusudhan 2019, Cridland et al. 2020), which has been previously measured with varying success using other ground and space-based facilities (e.g., Line et al. 2014, Brogi et al. 2014, Changeat 2022, Hoch et al. 2023). With access to the fundamental bands of molecules that contain carbon and oxygen in the mid-infrared, the fidelity of measurements of carbon and oxygen abundance should greatly improve. Beyond these tracers, species like nitrogen may also trace bulk formation processes, while less abundant refractory elements such as sulfur may inform pebble versus planetesimal accretion (e.g., Ohno & Ueda 2021, Turrini et al. 2021, Crossfield 2023). In the Solar System, isotopes such as deuterium and $^{13}C$ inform whether material was sourced from the interstellar medium or processed in the Solar System (e.g., Morley et al. 2019, Zhang et al. 2021, Gandhi et al. 2023).

One particularly powerful aspect of *JWST* is not only the ability to measure these abundances to high precision, but also the ability to probe planets across a variety of separations, ages, and masses. Large surveys of statistically meaningful swaths of the population may reveal abundance trends, such as patterns with age that may inform migration timescales or with mass that may correlate with formation mechanism. Observations of planetary disks are an important link between the phenomena of formation and the eventual properties of mature planets, including their atmospheres.

These science goals under this key science theme are closely aligned with the Astro2020 Decadal Survey Science Theme: Worlds and Suns in Context, which describes the importance of atmospheric characterization of diverse worlds. Broad and detailed atmospheric characterization is also listed in the Astro2020 Decadal Survey Priority Area: Pathways to Habitable Worlds, as it provides an understanding of the full exoplanet population and aids in the hunt for potential diagnostics of habitability.

Sample of White Papers associated with this Science Theme: 9, 10, 12, 14, 17, 27, 30, 32, 37

## Science Theme 3: Understand exoplanets in the context of their stellar environments

Understanding planets in the context of their stellar environments was a ubiquitous science theme across all planet types from the small terrestrial worlds, to the largest hot Jupiters. All responses that addressed this, acknowledged the unique need for *HST* to enable this science. These views are also deeply embedded within Science Themes 1 and 2. For example, the cosmic shoreline, featured in Science Theme 1, relies on understanding the cumulative XUV/UV stellar insolation. Similarly, in Science Theme 2, understanding photochemical processes across all planetary systems will also require an understanding of the host stars. Overall, we categorized these themes into three major topics:

### 3.1 Understanding how stellar heterogeneities affect observable planetary spectra

Stellar contamination from the Transit Light Source Effect is emerging as a major challenge for observations of transiting exoplanets. This effect is caused by the inhomogeneity of the stellar photosphere: the region of the star occulted by the transit chord may have a slightly different spectrum from the unocculted region, leading to spurious features in the exoplanet transmission spectrum (Rackham et al. 2017, 2018). Evidence for the Transit Light Source Effect has already been seen in several *JWST* transmission spectra for rocky planets (Moran et al. 2023, May et al. 2023, Kirk et al. 2024), and for the case of TRAPPIST-1b, there are spurious features from the star with amplitude ten times larger than the expected planet signal (Lim et al. 2023). Overall, there is consensus that understanding how stellar contamination affects the full range of spectra (from UV-optical to near-IR) will be an important science question that can be tackled with the strategic use of *JWST*, *HST*, and other facilities such as the *Pandora* SmallSat mission (Quintana et al. 2021).

### 3.2 Understanding how X-ray and ultraviolet emission sculpt planetary environments

Understanding the full spectral energy distribution of stellar spectra is of great astrophysical interest. Though the optical to infrared environments of stars can be generally understood from observations and theoretical investigations, the X-ray to UV environments are much more challenging to both observe and model (e.g., Linsky 2014). Yet, these short wavelengths are critically important in sculpting planetary environments, motivating previous treasury programs with *HST* (e.g., France et al. 2016). There is broad consensus that accurate measurements of these wavelengths are needed to assess the disequilibrium chemistry of gas giants (e.g., Zahnle et al. 2009). In particular, the first *JWST* transiting science observations for exoplanet WASP-39b foreshadowed how prevalent photochemistry would be in future observations (Tsai et al. 2023). Pushing toward smaller planets the stellar UV environments will be critical to accurately modeling (e.g., Teal et al. 2022) and interpreting detected features, especially those that are proposed biosignatures (e.g., Meadows 2018). Lastly, understanding how mass loss shapes the evolution of atmospheric environments is of prime focus for all planetary types and will be critical to understanding planetary populations.

### 3.3 Understanding how the magnitude of stellar activity influences atmospheres

A particular consideration for the stellar environments of M-dwarf-planets and the atmospheres of young planets is the magnitude of stellar activity (e.g., flaring). The understanding of stellar activity, across full diversity in stellar type was identified as a key science theme in many white papers and survey responses. Stellar flares, in general, have the potential to contribute a significant fraction of the total radiation budget of some planets. They also have the potential to contaminate the observations of spectra (e.g., Lim et al. 2023, Howard et al. 2023). Therefore, to accomplish both of the previously mentioned goals in Science Theme 3, it is important to also consider time-domain monitoring of host stars.

These three science goals within this Science Theme that emerged from the white papers and survey responses follow closely with the Astro2020 Decadal Survey Science Theme: Worlds and Suns in Context and the Priority Area: Pathways to Habitable Worlds. Tackling all three topics surrounding stellar environments would allow the community to understand how stellar activity can impact our planetary observations, identify the most extreme stars and stellar populations in the context of planetary hosts, understand how a planet's interaction with its host star and planetary system influence its atmospheric properties over all time scales, and identify what the range of potentially habitable environments around different types of stars may be.

### 3.4 Understanding planet occurrence as a function of stellar evolution

While the majority of exoplanet detection surveys for exoplanets are left to other facilities, *JWST* and *HST* potentially have unique roles to play in the discovery of new planetary systems. In particular, these telescopes offer the opportunity to explore planet occurrence as a function of stellar evolution, probing planets in extremely early phases of evolution using accretion indicators from *HST* (e.g., Zhou et al. 2021), or in the post-main sequence phase via sensitive imaging searches around white dwarfs. While some forming protoplanets are already known (e.g., Keppler et al. 2018), much about the persistence of planets in the latest phases of stellar evolution remains unconstrained, though theory predicts they may survive (e.g., Villaver & Livio 2007, Veras 2016). Additionally, the full census of planet occurrence evolution around <100 Myr old systems is not complete, with the outer regions of solar systems unexplored for planets like Saturn and Neptune. *JWST* is capable of identifying such systems in imaging surveys, and will be the only facility capable of doing so in the foreseeable future (e.g., Carter et al. 2023).

Sample of White Papers associated with this Science Theme: 3, 6, 10, 18, 33, 34, 39, 40, 42

# 5. Recommendations for Optimal Timing

The community provided a wide range of responses to the charge associated with optimal timing and mechanisms for enabling exoplanet observations. Here we present, first, a set of recommendations to STScI for implementation of the highest priority timing topics, and second, a set of suggested study items to STScI for topics that merit additional investigation to better understand how best to harness the combined power of *JWST* and *HST* observations for exoplanet science.

**Recommendation 1: The proprietary period should be tied to the final visit for multi-epoch observations.**

Many exoplanet transit, secondary eclipse, and direct imaging observations searching for small signals require multiple visits to achieve the required signal-to-noise ratio, astrometric motion, etc. Therefore, in order to achieve the science goals of the approved proposal, a complete set of visits is required. Additionally, many transiting planet observations take several months (up to multiple observatory cycles) to fully schedule. In order to enable the selected science investigation to complete their analysis before those data can be retrieved from the archive for similar or complementary science investigations, we recommend ‘starting the clock’ on the proprietary period with the execution of the last visit for a given target. In particular, we feel that this policy change would provide valuable protection for the data that is often used by early-career scientists who may be developing analysis capabilities for the first time as part of these programs. The WG recommends that the standard proprietary periods would remain in effect unless the proposer specifically requested this change and included a description of the science motivation in their proposal. Issues associated with proprietary periods have also been raised and discussed at the STUC and JSTUC meetings. This has led to a community survey on the topic which also identifies that this issue is critically important for exoplanet observers.

**Recommendation 2: Observing programs that may be impacted by programmatic timelines should be scheduled with high priority.**

The WG recommends that two types of observing programs, once selected by competitive peer review, be prioritized for scheduling. First, multi-cycle *JWST* exoplanet programs (e.g., transit observations of long period planets, direct imaging observations requiring long temporal baselines to measure orbital motion) should begin as soon as possible. These programs will span multiple cycles and are the most vulnerable to instrument or facility anomalies and/or degradation. Second, given the limited lifetime of *HST*, any DDT program with an *HST* component should execute as soon as possible.

**Study Item 1: The feasibility of simultaneous *JWST* and *HST* (UV) observations.**

The WG received several recommendations for simultaneous *JWST* and *HST* observing programs, however, the feasibility of true simultaneity between the observatories has not been fully explored. Given the experience of the existing *HST* joint programs (with X-ray or ground-based facilities), it is likely that there are practical limitations on visibility, sky coverage, etc that are not captured in APT. We suggest that STScI should provide some high-level guidance to proposers on what could be expected from these observations.

**Study Item 2: Forward funding of observing programs to support analysis and model preparation, and personnel acquisition necessary to carry out the proposed work.**

The WG considered numerous community inputs on the "effective duration" of an *HST* or *JWST* observing grant being significantly less than the nominal 3 years. Often, the academic hiring cycle can have a 6–12 month lag time. If for instance, a program requires hiring a postdoc or graduate student to complete key tasks, this can reduce the effective duration of an analysis and publication grant to ~2 years, sufficient to complete the analysis but often insufficient to complete associated modeling and publication. This is particularly true for time-constrained observations (like exoplanet transits) where a full set of visits may take one or more cycles to complete. Similar arguments can be made for being able to support existing personnel to develop analysis software or models specific to a given observing program. The WG suggests that STScI investigate the possibility of enabling GO awardees to receive sufficient pre-observation funding to enable project-critical personnel and tools to be in place prior to the first visit of a program.

# 6. Recommendations for Scale of Resources

The WG considered community feedback on the appropriate scale of resources to support exoplanet science with *JWST* and *HST* and makes four recommendations and identifies one Study Item:

**Recommendation 1: Provide dedicated and robust funding support for DDT.**

To reap the full benefits of data taken with new facilities, like *JWST,* requires significant funding. It is required to support the holistic needs of researchers, including model development and computing, lab measurements to support calibration of both data and models, and the development of customized data reduction/modeling software. Scaling from the typical funding of ~$10K/hr in the early cycles of JWST, the DDT project proposed by this working group will

require at least ~$10K/hr * 500 hr = $5M to be sufficiently funded. It is important that at least a significant portion of the necessary funding is received with the initial time allocation. Lessons from the *HST* ULYSSES large program (~1000 hours, the largest program to date with either *HST* or *JWST*) indicate that receiving additional support via the archival proposal route was difficult, with respect to the timing and organization of the archival proposal. We received strong community concern regarding underfunding of *JWST* programs, based on the ERS experience. Without sufficient and timely funding, we will not see appropriate science returns from the valuable data of the DDT program. We may also expect inequality in who is able to analyze the data, with those at more well resourced institutions having a clear advantage.

**Study Item 1: Explore providing inclusive, program management support.**

The community generally appreciated the aspirational goals of inclusiveness and data analysis support available through the two PI-driven ERS exoplanet teams, but expressed concerns that this relied strongly upon the self-organization and efforts of individuals in the community (often unfunded). We recommend that STScI explore ways to provide some degree of support, instead of leaving it to highly resourced individuals / groups from the community, for both the proposed DDT program stemming from these efforts and for exoplanet science in general (e.g., the $10^4$ Hour Exoplanet Survey). This could be as informal as purchasing and setting up a Slack space for the community, establishing a working group leadership model, organizing community-building data-reduction, analysis, and modeling meetings, or having a contact scientist at STScI take on a quasi-project management role.

**Recommendation 2: Prioritize supporting *HST*, ground-based, and X-ray observations to enable holistic exoplanet studies.**

While *JWST* observations have been transformative for the exoplanet field, *HST*, ground-based, and X-ray observations are synergistic with *JWST* and are critical for understanding phenomena important to exoplanets. *HST* UV and X-ray observations are a key resource for understanding the presence or absence of an atmosphere on rocky planets and photochemistry on all types of planets. In general, there are many exoplanet projects that require a holistic approach to characterizing the exoplanet and its stellar radiation environment. Thus, to maximize the scientific return with *JWST* via complementary observations with other facilities, we recommend ensuring support for *JWST* and *HST* simultaneous observations and coordinated observations with facilities across all wavelengths, and enabling NASA XRP support for a larger range of ground-based observations.

**Recommendation 3: High demand/need in exoplanets supports allocating more than 500 hrs of DDT time.**

In the early *JWST* cycles, roughly one third of all *JWST* time has been allocated to exoplanet science. Over the anticipated lifetime of *JWST*, this will yield $>10^4$ hours dedicated to exoplanet observations. While the suite of accepted GO programs will be heterogeneous, there will be significant added value in collating individual GO programs to produce essentially a serendipitous survey of exoplanets over a wide range of exoplanet masses and irradiation levels. We discuss this outcome in Sec. 2.3 as the $10^4$ hr *JWST* Exoplanet Survey and provide more context for strategic implementation in Sec. 8.2. We recommend providing support for observations that will provide archival/legacy value in the context of a broad $10^4$ hr Exoplanet Survey. Our recommended 500 hr DDT Concept is meant to kickstart this serendipitous large survey, but future DDT programs with similarly large time allocations would aid in producing large datasets that are more uniform than can be expected via a conglomeration of data acquired in single- or few-object GO programs. In particular, our second-place DDT concept (G395H survey of a wide range of exoplanets), if realized as either a set of large programs or as a future DDT program, would be an important step toward achieving this goal producing a legacy dataset associated with a $10^4$ hr Exoplanet Survey.

**Recommendation 4: Provide the most advanced data products possible, a necessary resource for projects with dynamical scheduling / hierarchical implementation.**

Based on the community feedback, participants in both the Transit Exoplanet and Direct Imaging Exoplanet ERS teams relayed that significant amounts of time and effort went into data reduction and analysis for these programs, which is to be expected at the start of a new mission. However, now that the observatory is coming into a more mature stage of observations, to more optimally utilize *JWST* data requires that the community has access to advanced, well-calibrated data products. In particular, having access to advanced data products soon after observations are taken is critical for projects which are dynamically scheduled or implemented on a step-by-step basis. For example, programs that observe multiple transits in order to build up a sufficiently high S/N ratio lightcurve currently estimate S/N ratio from tools such as the ETC, then request a specific number of visits to reach that S/N ratio, and only determine after the fact whether the required S/N ratio was met. Advanced data products available soon after an observation would enable such a team to dynamically determine the S/N ratio in their stacked lightcurves, and better plan subsequent observations. The recommended DDT Concept, which itself requires large numbers of visits for some targets and would benefit from dynamical scheduling, would be a great place to develop these tools with the goal to eventually make them available to all programs. Examples of similar advanced data products include the Hubble Advanced Spectral Product (HASP; Debes et al. 2024) and those coming out of ULLYSES.

# 7. Recommendations for a DDT Concept: A Survey of Rocky Worlds

**Recommendation 1: Use 500 hours of *JWST* time and 240 orbits of *HST* to perform a survey of rocky worlds to measure the cosmic shoreline and identify exciting targets for follow-up observations.**

One of the common high-priority science questions that arose in the community feedback was the nature of small planet atmospheres (see Sec. 4 on Recommendation for Key Science Themes). For rocky planets, the most fundamental question is whether an atmosphere is present. If yes, what is the chemical composition? Which planets might have the right ingredients for life to arise? These questions are challenging to address *a priori* because of the wide range of physical processes that shape secondary atmospheres – everything from the initial volatile inventory, to the outgassing history and atmospheric escape processes, and even the presence of life. Fortunately, thanks to *JWST* it is now possible to address these questions observationally for the first time.

*JWST* provides a unique new opportunity to characterize the atmospheres of rocky planets orbiting small, M-dwarf stars. Scientifically, these planets are particularly interesting for follow-up because they are the most numerous rocky planets in the galaxy. Practically speaking, M-dwarfs are the most abundant stellar type, and they also host more rocky planets than Sun-like stars, on orbits within 100 days. Moreover, *JWST* has the unique ability to characterize the atmospheres of planets in these systems with transit and eclipse observations. M-dwarf hosts are essential: the smaller, cooler star makes transits more likely, more frequent, and deeper, for a given planet size and temperature. Together, these factors are colloquially known as the "M-dwarf opportunity".

*JWST* observations are the only avenue to characterize the atmospheres of a representative sample (~10s) of these planets in the foreseeable future. The observations require a (1) large mirror to detect the small signals from the atmosphere; and (2) infrared wavelength coverage with access to the rich array of molecular features between 2–12 micron and sensitivity to thermal emission for temperate planets. ELT-scale ground-based observations may give access to the atmospheres of a handful of rocky worlds, but this hinges on removing telluric effects at a level of precision orders of magnitude better than current capabilities (e.g., López-Morales et al. 2019, Palle et al. 2023). Looking farther ahead, not even *HWO* will be capable of this science. *HWO* will be optimized for directly imaging Earth-like planets orbiting Sun-like stars, and will not have access to large numbers of M-dwarf systems, due to the close inner working angle between temperate planets and small stars.

*JWST* has already begun to characterize rocky planets orbiting M-dwarfs, see Figure 6. The observational approach is to measure transits, secondary eclipses, or even full-orbit phase curves for some systems. From these early results, thermal emission measurements are emerging as a powerful tool to explore whether an atmosphere is present or not. The key idea is that thick atmospheres efficiently recirculate heat from the dayside to the nightside of tidally locked planets (expected for the short-period worlds with periods shorter than ~10 days, extending to the habitable zones of late-type stars; Kasting et al. 1993). To zeroth order, cooler dayside temperatures imply thicker atmospheres with higher surface pressure (Seager & Deming 2009, Selsis et al. 2011, Koll et al. 2019). Notably, secondary eclipse observations are not affected by the Transit Light Source Effect, because the full disk of the star is observable during the entire eclipse. Given the complexity in interpreting transit observations in the face of stellar contamination, the WG recommends prioritizing secondary eclipse observations to kickstart our understanding of the prevalence and diversity of rocky planet atmospheres.

Figure 6 shows targets relative to hypothetical cosmic shorelines. The shoreline is an empirical relationship inspired by the Solar System that separates bodies more likely to have atmospheres from those less likely to have them (Zahnle & Catling 2017). Higher escape velocities and lower insolation (particularly UV insolation) are more conducive to the survival of an atmosphere. The survivability of rocky planet atmospheres is likely to be driven by XUV emission for M dwarf hosts. The importance of UV in driving atmospheric escape highlights the need for *HST* observations to complement the *JWST* DDT. Auxiliary observations in the X-ray with *Chandra* would also be valuable.

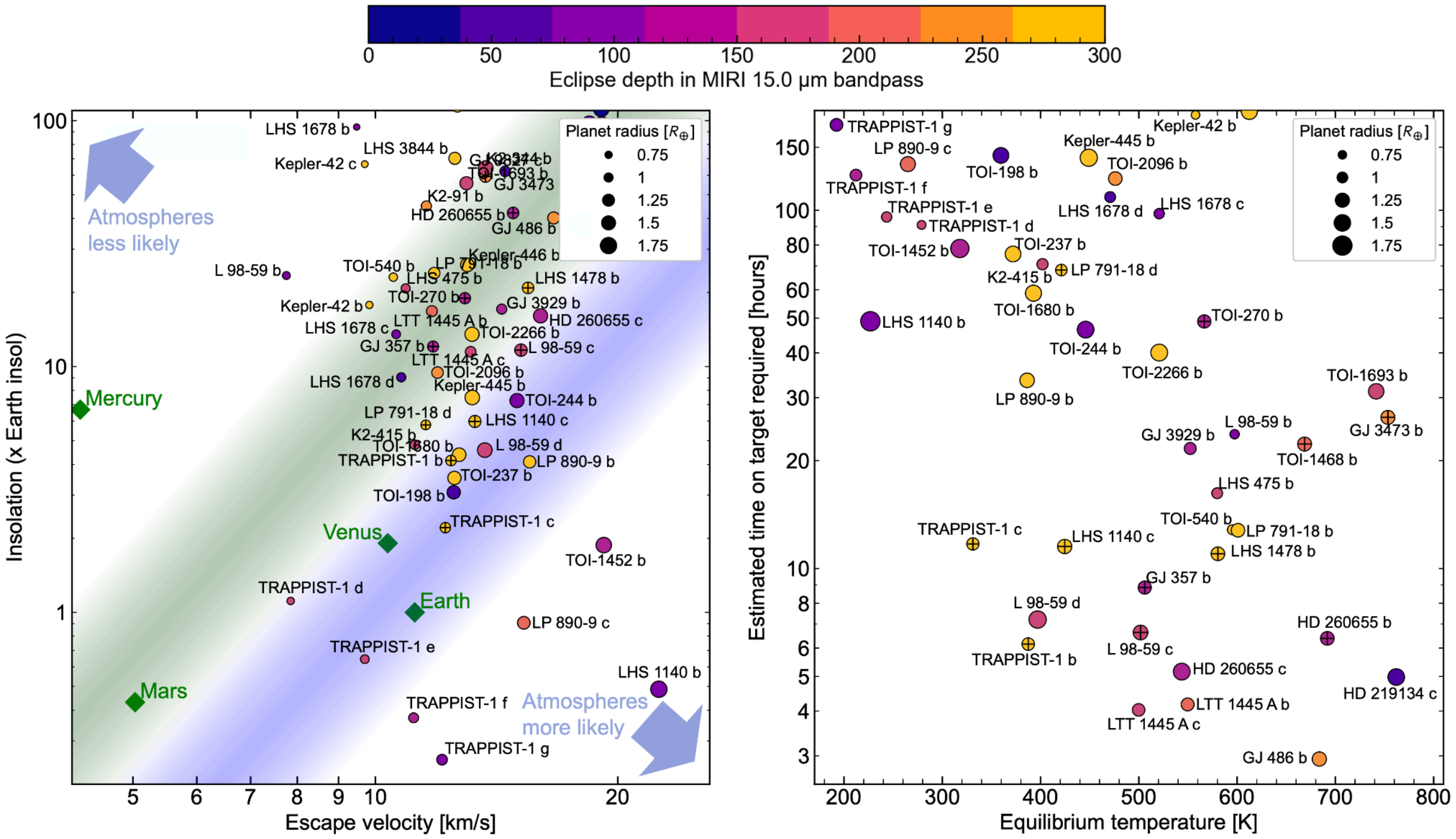

**Figure 6:** (***left***) Potential targets for the rocky worlds survey compared Solar System planets and the cosmic shoreline. The objective of this DDT concept is to identify the location of the cosmic shoreline for M-dwarf rocky worlds, with two hypothetical cosmic shorelines indicated by the green and blue fuzzy lines. The marker size indicates the planet radius, and the marker color indicates the expected eclipse depth at 15 µm, i.e., the absolute strength of the signal to be observed. (***right***) The equilibrium temperature of potential target planets (horizontal axis) versus the approximate amount of JWST time required to distinguish between the bare black rock scenario and scenarios with full heat redistribution at >99.7% probability (vertical axis). Approximately half of the planets indicated could be observed in the recommended DDT program. Planets with at least one 15 µm eclipse observed or scheduled are marked with a black cross.

The core idea for the DDT Concept is to observe secondary eclipses of a sample of 15–20 rocky planets orbiting M-dwarfs. The goal of the observations is to measure the dayside temperature as a proxy for the planet's heat redistribution, with the objective of distinguishing between full versus zero heat redistribution at 5σ confidence (assuming a gray atmosphere). Full heat redistribution is expected for high atmospheric surface pressures (10 bar), whereas no redistribution is expected for airless planets. A detailed implementation checklist is provided in Section 8.1.

With an allocation of 500 hours, it will be possible to roughly double the sample of rocky planets with thermal emission measurements, and push to significantly cooler temperatures (where atmospheres are more likely). The planets should have a range of equilibrium temperatures and escape velocities, ideally spanning the cosmic shoreline. The sample should also cover a range of host star spectral types across the red dwarf mass sequence. We discuss the trade-offs between a survey approach versus intensive observation of one system (e.g., TRAPPIST-1) in the Frequently Asked Questions below.

The *JWST* DDT should be combined with *HST* FUV and NUV observations to characterize the host star. The stellar UV spectrum is an essential input for models of atmospheric loss and photochemistry, and must be observed now while *HST's* unique UV capability is still available.

**Target Selection and Observing Strategy**

Based on a preliminary assessment from the WG, a total of 15–20 rocky planets could be included in a 500 hr DDT program, with equilibrium temperatures spanning the range 200–600 Kelvin. The expectation is that the time allocation could range from 2–15 eclipses per target, with roughly 5 hours per visit. The coolest targets are expensive (estimated to take 50–100 hrs), but may also be more likely to have atmospheres and should take precedence over a larger number of hotter targets. Ideally the targets should span the cosmic shoreline: a diversity of temperatures, host star types, and planet masses should be prioritized.

Secondary eclipses of the planets should be observed with *JWST*/MIRI photometry using the 15 micron F1500W filter, identical to the observing strategy for TRAPPIST-1b and c (Greene et al. 2023, Zieba et al. 2023). The 15 micron filter covers an absorption band of carbon dioxide, which is expected to be both one of the strongest absorbers and one of the likeliest atmospheric constituents for rocky planets over a wide range of temperatures (Lichtenberg & Clement 2022). The signal-to-noise for thermal emission measurements is also typically near maximum at this long wavelength for the relatively cool planets to be targeted with DDT observations.

**Frequently Asked Questions**

- Why not exclusively study TRAPPIST-1?

  Several white papers and survey responses advocated for intensive study of the TRAPPIST-1 system alone. TRAPPIST-1 is indeed unique in the number of rocky worlds accessible for study; however, the WG recommendation is to prioritize a diversity of systems, rather than putting all eggs in one basket. The TRAPPIST-1 planets are also already under intensive scrutiny in JWST Cycles 1 and 2, with nearly 40 unique visits covering all 7 planets. Further additional deep observations were recently approved in Cycle 3. These initial observations will provide an important foundation for guiding the design of future observations.

- Why not observe each target at multiple wavelengths?

  The WG weighed the tradeoff between broader wavelength coverage and number of targets, and opted to prioritize a larger sample. Even with a single wavelength band, it is possible to search for trends in heat redistribution over the population of rocky planets. Outliers or other interesting targets will be prime candidates for follow-up at additional wavelengths, opening the possibility for community-led GO programs. With this additional wavelength coverage for individual targets, it will be possible to break degeneracies between heat redistribution and vertical temperature structure that could arise if an atmosphere is present.

- Why not transits?

  The nature of rocky planet atmospheres is still almost completely unexplored. It is not known which planets have atmospheres, or what the chemical composition may be of any atmospheres that are present. Clouds and hazes are an additional unknown, though extrapolating from the Solar System and gaseous exoplanets, their presence would not come as a surprise. Given these multiple large uncertainties, informative transit observations are difficult to design. The requisite S/N is poorly constrained due to the unknown mean molecular weight, and a flat spectrum can be due to either an airless planet or high-altitude clouds and hazes (e.g., Kreidberg et al. 2014). Beyond this, stellar contamination is emerging as a serious issue, and effective mitigation will require major improvements in theoretical models of stellar photospheres and/or different techniques and strategies (either observational, theoretical, or both) to the ones commonly employed in transmission spectroscopy studies (Lim et al. 2023, Rackham et al. 2023). By contrast, secondary eclipse observations of dayside thermal emission are a more straightforward route to assessing first and foremost whether an atmosphere is present. Any atmosphere will tend to redistribute heat, regardless of chemical composition. Further, stellar contamination is not an issue for thermal emission spectra. The reason is that in transit, the average stellar spectrum in the transit chord can differ from the rest of the photosphere, leading to spurious spectral features in transmission (Rackham et al. 2018). This effect does not apply to emission spectroscopy, where the planet passes behind the star and therefore the same region of the star is always visible during the observation.

**Legacy Value and Appropriateness for DDT**

In sum, the proposed DDT program is a high-risk, high-reward program that will answer a key science question outlined in the Astro2020 Decadal Survey Priority Area: Pathway to Habitability. It will definitively identify which M-dwarf rocky planets have atmospheres, which *JWST* alone can do. It will also enable efficient follow-up of any planets that do have atmospheres, to determine their chemical composition and surface pressure. It will push the “existence test” for atmospheres all the way down to the habitable zone, providing an important stepping stone on the path to habitability. The results from this program will inform the design of both future *JWST* observations and future facilities like *HWO*. There is strong synergy between the proposed program and *HST*/UV observations of the host stars, which is urgently needed while *HST* is still operational.

A DDT Program is needed to obtain systematic demographic study of the rocky planet population. Building such a sample through the normal GO process is a challenge because each additional planet must be separately justified. In addition, the DDT route will also give access to the coolest targets (below 400 K). These targets are perhaps the most likely to have atmospheres, but also the most expensive (up to 100 hours required!), and unlikely to be accepted through GO proposals. An additional benefit of a DDT program is dynamic scheduling; for some targets, the exact eclipse time is not perfectly known due to orbital eccentricity. In these cases, the optimal observing strategy is to start with a large window to catch the eclipse and then narrow it after the eclipse time is well-determined. It is also possible that the secondary eclipse could be missed entirely; statistically this is very unlikely to happen for many planets, but the value of a large DDT program is that the overall success does not hinge on any single planet in the sample.

Finally, the program does carry scientific risk, in the sense that one possible outcome is that none of the planets have atmospheres. Similar to the Hubble Deep Field observations, when it was unknown if any galaxies were present at high redshift, here we do not know how common atmospheres are on rocky planets. We will never know unless we look – and with a large, focused effort through this DDT concept, we will make a transformative step forward in answering this question.

Sample of White Papers associated with this DDT Concept: 15, 16, 21, 25, 26, 28, 29, 36, 38

# 8. Implementation, Data Products and STScI Support

## 8.1 Support for the Recommended DDT Concept

The main objective of the recommended DDT Concept will be to determine the prevalence of atmospheres on rocky planets orbiting low-mass stars. Planets with low stellar irradiation are believed to be more likely to have retained their atmospheres. We therefore regard it as particularly important to extend the current sample of rocky planets with *JWST* eclipse observations to the planets with low equilibrium temperatures, pushing the limit of what can be achieved with *JWST* in a large 500 hour DDT program.

As a general guideline, we therefore recommend a target selection that dedicates a large fraction of the assigned telescope time (of the order of 40–70%) to the most intriguing cool rocky planets between 200 and 450 K suitable for thermal emission observations. Beyond that, we recommend that the program also includes warmer planets up to an equilibrium temperature of approximately 600 K as long as the planet's escape velocity is still high enough to plausibly retain an atmosphere against thermal evaporation. These warmer planets are comparatively less expensive in terms of telescope time and a larger number of them can be probed within a relatively small fraction of the 500 hour DDT program (Figure 6). Overall, we recommend a target selection that samples the temperature range between approximately 200 K and 600 K as well as possible within the time constraints of the program.

We also recommend including targets across many different planetary systems rather than focusing only on a small number of individual planetary systems. The planetary atmospheres orbiting a given host star may have undergone the same fate due to a host star's potential history of high stellar activity. At the same time, including two or more planets orbiting the same star for a couple of systems can be helpful in separating the effects of stellar activity and planet temperature, thereby enabling a comparative exoplanetology study.

Figure 6 displays potentially suitable targets for the *JWST* DDT rocky worlds eclipse survey. The approximate telescope time required for each target is shown across a range of zero-albedo equilibrium temperatures. For any given target, the implementation team should consider both the time required for secondary eclipse detection in the limit of photon noise as well as the expected absolute strength of the eclipse signal to be observed, that is, the eclipse depth in the MIRI 15 μm bandpass assuming a zero-albedo bare rock (the color shading). The latter is important because the survey will likely aim into unchartered territory in terms of MIRI eclipse depth precision and systematic noise may set limits on the absolute eclipse depth precision achievable. The observations should be defined to obtain a 3σ confidence between the limiting case scenarios for the temperature with full versus zero heat redistribution. Each of these targets needs to be carefully vetted to assess their suitability for the *JWST* DDT rocky worlds eclipse survey before becoming part of the program. We include a checklist below.

Importantly, the implementation team should engage with the scientific community and have representation from the community. An online tool should be made available for the scientific community to point out potential challenges, show-stoppers, or special opportunities for any given target. The objective is for the Implementation Team to make the most informed decision possible in the target selection process. For that, the implementation team should make the target list publicly available as soon as possible after an initial target selection. A procedure for dynamical scheduling should be developed. This would involve evaluating the early individual observations to make sure they are as expected and having a policy for adjusting targets or observing parameters as needed.

We also note that quantitative assessments of the host stars' UV emission spectra and stellar-activity levels will be essential in interpreting the *JWST* observations. We therefore recommend that the implementation team performs contemporaneous UV observations with *HST* for all targets in the *JWST* rocky worlds DDT survey. The implementation team should also encourage the community to prioritize the selected targets for radial-velocity mass characterization and additional contemporaneous host star observations, e.g., *Chandra*, *XMM*, and ground-based observations. Community engagement efforts should be implemented to help coordinate this community effort.

**Implementation checklist**

Here we provide a checklist of important factors for the implementation team and STScI to consider:

1. Targets should be rocky planets, i.e., those lacking a H/He envelope. We recommend selecting targets below a threshold radius of ~1.75 Earth radii (Rogers 2015). Larger planets could potentially be included if they have a precise (5σ) measured mass and radius consistent with an Earth-like interior composition.
2. The targets should be confirmed planets, ideally published in peer-reviewed journals. Due to the fast-paced nature of exoplanet discovery, the implementation team could also consider targets submitted for peer review but not yet published. Mass measurements are desired; if not yet available, the implementation team should confirm that a mass measurement is feasible with realistic instruments in the near future and communicate to the community that such observations would be of high value.
3. The number of eclipses per target should be chosen such that the temperature measurement can significantly distinguish (3σ confidence) between limiting case scenarios for the temperature (full versus zero heat redistribution).
4. The secondary eclipse must be visible during the observations. This means:
    a. The secondary eclipse occurs within the timing window. Secondary eclipse times can span a wide range for targets with non-zero eccentricity (Alonso 2018). To account for this, the implementation team should prioritize planets with a short tidal circularization timescale, or targets in multi-planet systems (which are known to have low average eccentricities; Van Eylen et al. 2019). For some targets, it may be prudent to add baseline to the first eclipse observation to pin down the eclipse ephemeris.
    b. The secondary eclipse probability is above 99%. For targets with high impact parameters, the planet is not guaranteed to be eclipsed by the star along our line of sight. The implementation team should verify that a secondary eclipse is likely based on the inferred orbital parameters and their uncertainties.
5. To the extent possible, the implementation team should aim for a diverse selection of planet equilibrium temperatures, escape velocities, and host star spectral types across the M-dwarf spectral sequence, to explore the cosmic shoreline idea outlined by Zahnle & Catling (2017) (see Fig. 6). To achieve balance across equilibrium temperature, approximately half (40–70%) of the time should go to cooler objects, while a larger number of warmer objects can be obtained in a comparable amount of time.
6. The targets should be cool and massive enough that retaining some atmosphere is within reason. As a strict limit, the thermal escape velocity for $CO_2$ should be no more than 1/6th of the escape velocity. This requires that $T*R_p/M_p < 10^4$ (where T is planet equilibrium temperature in Kelvin, $R_p$ is the planet radius in Earth radii, and $M_p$ is the planet mass in Earth masses).

7. STScI should provide as much scheduling flexibility as possible for the implementation team. A dynamical scheduling procedure should be outlined so that the team can adjust the target list and observing strategy after evaluation of the initial observations.
8. An up-to-date target list for the DDT program should be made publicly available to facilitate follow-up efforts and coordination with the broader exoplanet community.
9. To ensure the most efficient use of the DDT allocation, STScI should promote as much community engagement and communication as possible in terms of target vetting, auxiliary observation gathering, data analysis and interpretation. Ideally, a forum can be promoted where efforts by community members can be communicated and opportunities for collaboration can be facilitated.
10. *HST* FUV and NUV spectroscopic characterization should be started as soon as possible to ensure that comprehensive SEDs are available for each *JWST* DDT planet host. While contemporaneous measurements would be desirable, this places unrealistic scheduling requirements and *HST* orbit allocations given the large number of total eclipse measurements. Building on existing M dwarf studies that have been used to develop widely used stellar inputs (France et al. 2016), we suggest 15 orbits per star per campaign. This conservative estimate is broken into 5 orbits for COS G130M to monitor a range of stellar emission lines for flares, 3 orbits for COS G160M to complete the FUV characterization, 4 orbits for STIS G140M for Lyman-alpha reconstruction, and 3 orbits for STIS G230L for the NUV spectrum. 15 orbits per star for an approximate 16 total planet hosts drives the total recommended UV spectroscopic allocation to ~240 orbits.

**Additional considerations:**

- Meta-data products will be important, e.g., orbital and planetary properties with references. Links to SIMBAD and NASA Exoplanet Archive may be helpful for all targets in order to connect with other repositories of data on these targets (e.g., RVs, transit timing, etc).
- Data products of the individual observations will need to be organized. Note for many targets several eclipses are required. Previously observed targets will also satisfy the criteria, and they should be listed with proper acknowledgment of proposers and publications.
- Fast, high-level data products, similar to ULLYSES and HASP, should be made available as soon as possible and used to optimize the dynamical scheduling of further visits.
- Recent experience should be taken into account with the active M stars and the *HST* flare safety protocols.
- Among the risks of these observations are the uncertainty in the planetary mass (for characterizing the atmosphere) and the orbital eccentricity (for the timing of the secondary eclipse). Auxiliary observations to more precisely measure these quantities should be prioritized. Also, contingency in the observational design should be considered to accommodate uncertainty in these parameters.
- A STScI website, similar to ULYSSES, should be made to organize the data and meta-data products. It can also serve as a focal point for community engagement.

## 8.2 Support for $10^4$ Hour Exoplanet Survey

In response to the strong community support for broad studies of various exoplanet populations, and the inevitable accumulation and synergy of GO-driven projects, in what we call the $10^4$ hr Exoplanet Survey, we feel it is important to advocate for support for such an effort. We have argued that this kind of strategic planning will pay large dividends in the eventual transformative JWST dataset of exoplanet observations. In this section, we make recommendations for supporting such an endeavor, as well as sketch out the details of an alternative DDT Concept that was also considered by the WG, which would act to kickstart the creation of a strategic exoplanet population survey.

**Recommendation 1: Provide support for observations that will provide archival and legacy value in the context of a broad $10^4$ hr exoplanet atmospheric survey in the form of an exoplanet demographic observation website and incentives for programs that qualify.**

We advocate for an actively maintained STScI demographic observation website to optimize exoplanet observations and support a community-driven $10^4$ hour Exoplanet Survey. This would be separate from the DDT Concept study, and act as a resource and stimulus for community engagement in developing and augmenting individual programs that could also serve as a vital part of a population-wide survey. This investment would have significant returns by optimizing the eventual *JWST* exoplanet archive. With the current resources, it is very difficult and time consuming for GO proposers to identify which exoplanet targets have been archived or scheduled and in which geometries (transit, eclipse, and phase curves). Confounding factors such as inconsistent naming schemes, differing ephemeris definitions, withdrawn and failed observations make the task non-trivial.

The demographic website would provide a growing database of observations of populations across various properties (e.g., mass, temperature, metallicity, age, etc). This would provide a scientifically motivated entry point into the *JWST* archive. It would identify gaps in targets or wavelength coverage. It could also be a place for community members to identify interest, efforts, or resources. Such an exoplanet demographic observation website would be also generally useful for proposal planning and evaluation, as duplicate observations could be efficiently and robustly evaluated. Various versions of websites or lists have been generated by individuals or small groups (e.g., TrExoLiSTS: Transiting Exoplanets List of Space Telescope Spectroscopy; Nikolov et al. 2022). This demonstrates the high value and interest of such a resource for the field. However, they often suffer from limited scope or a failure to stay up-to-date. A resource that includes system parameters, integration with exo.MAST, and on-the-fly plots would be a powerful tool for the transit and direct imaging exoplanet communities.

We recommend considering further mechanisms to incentivize and coordinate GO programs over the coming decades toward a $10^4$ hour Exoplanet Survey. *HST* has used check-box initiatives such as the UV initiative, *TESS* initiative, and *JWST* initiative to incentivize strategic priorities in the TAC process. A similar strategic priority is recommended for the $10^4$ hour Exoplanet Survey. Further incentives such as time subsidies, as has been done to incentivize medium level proposals, or other TAC mechanisms should be studied.

### 8.2.1 A Giant Planet Survey

A compelling concept that was considered in depth was a comprehensive giant planet survey (GPS) using NIRSpec G395H observations which could be executed as a Multi-Cycle Treasury Program. The GPS would act to kickstart the larger $10^4$ hour Exoplanet Survey, as a broad range of exoplanets and related sub-stellar objects would be systematically and strategically surveyed. The GPS concept would deliver a statistical survey of giant planets across a wide range of the two key physical parameters of mass and temperature (see Figure 7). The GPS survey would span planet masses from Neptune though Jupiter up through brown dwarf masses, at temperatures from 500 to 2500 K. Such a large comprehensive planet survey has never been accomplished and would enable a broad range of science uniting three large communities: the transit, direct imaging, and brown dwarf communities. While a wide range of science would be enabled, both at the individual planet and population level, the GPS would deliver on two main science themes: planet formation and atmospheric chemistry. With a statistical survey, comparison studies extending from planetary to stellar regimes on these topics could reveal the larger-scale physical processes and trends.

**Planet Formation**

By targeting low mass transiting planets (0.3 to 1 $M_{Jup}$) and comparing them with more massive directly imaged planets and brown dwarfs, the metallicities and detailed abundances derived from *JWST* spectra could help answer longstanding questions about the differences in how planets and stars form. Brown dwarfs and perhaps directly imaged planets are thought to form like stars from gravitational instability, while Neptune to Jupiter mass planets form from the core accretion mechanism. Differences in the measured carbon-to-oxygen ratio and metallicity (C/O, *Z*) could indicate which objects form from which mechanism, and the transition between the two could be ascertained for the first time. *JWST* can uniquely probe oxygen and carbon species with *JWST*/NIRspec G395H, and is the only telescope and instrument combination capable of delivering spectra of both transit and directly imaged spectra (Ruffio et al. 2023).

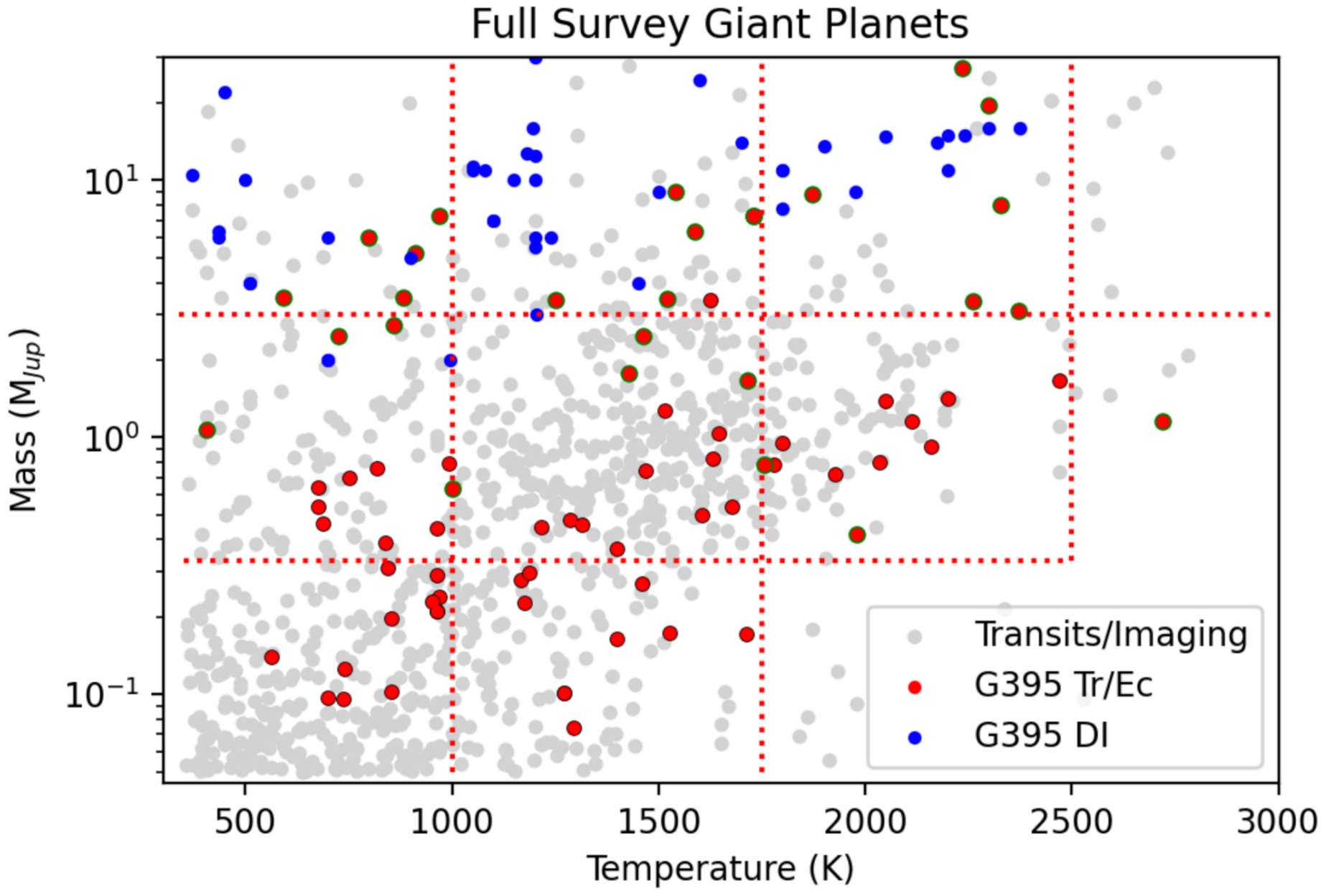

**Figure 7**. Mass and temperatures of known planets (gray) along with possible survey targets (blue and red). Red dotted lines depict bins in mass and temperature for statistical comparison. 50+ new targets, together with previously observed targets would combine into a ~100 planet survey. Sufficient statistics would be available (~10 targets/bin) to measure the mean <Z> and <C/O> as well as the astrophysical dispersion of these parameters to <25%.

**Atmospheric Chemistry**

In addition to planet formation, intimately tied to a planet's mass, a comprehensive survey across temperature would enable a range of science topics regarding planetary chemistry and atmospheric dynamics. To date, the available spectra for low-mass giant planets have largely constrained oxygen-bearing species, with *JWST* now accessing carbon and sulfur-bearing molecules. A range of science has opened up including: disequilibrium chemistry, mixing, photochemistry, atmospheric chemistry and dynamics in high/low metallicity and C/O regimes, cloud composition, photochemically generated hazes, spatial inhomogeneities, morning/evening terminator limb asymmetries, dynamics, and intrinsic variability. With a large statistical survey, sub-populations could be chemically identified (e.g., high or low Z planets), and importantly the astrophysical dispersion in key parameters of Z and C/O could be directly measured. This would have significant implications for future surveys, as the target numbers could be set ahead of time to constrain trends such as the mass-metallicity relation.

**Survey Scope**

*JWST* NIRSpec/G395H provides a rich spectral region which is optimal to cover the most important carbon and oxygen bearing molecular species (e.g. $H_2O$, CO, $CO_2$, $CH_4$) within a single spectral mode. Further species such sulfur can be measured as well. While other *JWST* modes could extend the wavelength coverage, a large survey with many targets covered statistically and uniformly with the same spectral mode was found best for a first-generation giant planet statistical survey. From *JWST* Cycles 1 and 2, about 27 targets have been completed which could be used as part of the survey. That represents about 41% of the needed observations. The TAC allocation rate from the Cycles 1 and 2 for observing giant transiting planets was ~7/year. At that rate, a full survey would not be completed for most of *JWST's* mission lifetime and target coverage and number in mass and temperature would be unlikely to enable a full statistical analysis. Large programs could accelerate this progress. However, the transit, direct imaging and brown dwarf communities are distinct and survey requirements are large, making it difficult to propose a similar GPS in the current normal GO process. With a single Multi-Cycle Treasury program, 500 hours could enable a statistical survey conducted in a uniform manner not achievable through the current GO program. We estimate a GPS would need about 350 hours to complete the transiting planets sample, which covers 40 planets, and approximately 150 hours would be needed to cover 16 directly imaged/brown dwarf targets.

**Risks and Rewards**

A comprehensive giant survey carries an inherently low observational risk. The methods are now proven and the signal-to-noise ratio for all of the targets would be large. For a given science topic, a risk would be that the final survey statistics may still be insufficient to constrain formation or chemistry models, if a large spread in the underlying property was discovered in the planet population. However, in this outcome the diversity in exoplanet properties would be statistically measured, and the necessary survey target numbers would be known for future follow-up efforts.

A large giant planet survey would provide a powerful legacy dataset for *JWST*. A plethora of discoveries and topics can easily be envisioned spanning the 3 distinct sub-fields: brown dwarfs, transiting exoplanets and directly imaged exoplanets. In addition, a GPS would provide a unique synergy between spectra for both transit and directly imaged planets obtained with one instrument for the first time, designed with statistical comparisons in mind from the start. The resulting rich spectral dataset could enable cross-disciplinary science where the whole survey would be much greater than the sum of its individual parts. The survey would also help motivate future observations, as observations at other wavelengths (e.g., SOSS, MIRI) could enable further science. Such a GPS would greatly benefit from *HST's* UV capabilities. A strong science motivation for observing transiting planets would be to study atmospheric photochemically generated molecules. UV spectra of their host stars is a critical input for the photochemical models, yet is very difficult to predict and requires observations of the host stars to understand the XUV environment of the planetary atmosphere.

**Recommendation 2: Enable a mechanism for Multi-Cycle Treasury Programs which could support large-scale exoplanet surveys.**

Sample of White Papers associated with this DDT Concept: 1, 2, 7, 9, 11, 13, 22, 27, 30, 32

# 9. Community Role

This report and the recommendations within it have been strongly shaped by the exoplanet community. As discussed above, we received considerable direct input from the community through the town halls, survey responses, and white papers. We also have made use of the Astro2020 Decadal Survey, which synthesized more generally the state of exoplanet science and highlighted important avenues for future research.

One notable feature of how the exoplanet community has thus far engaged with *JWST* data is the increasingly cohesive collaborations between different communities (stellar, exoplanet, and planetary) and experts in different techniques (e.g., data reduction/analysis, modeling/theory, spectroscopic lab support, ancillary observations at different wavelengths). For example, the revolutionary advance in data quality from *JWST* has often necessitated more direct accounting of the role of the star, more focus on the precision of line list for various opacity sources, and the use of more detailed, complex models in the data interpretation. We expect the prevalence of these cross-technique and cross-disciplinary collaborations to continue to grow as exoplanet science with *JWST* advances, enhancing the overall science from this mission. Both our recommended DDT Concept for a Survey of Rocky Worlds and our vision for a $10^4$ hour Exoplanet Survey provide exciting opportunities to knit together more subfields.

The Survey of Rocky Worlds will revolutionize our understanding of both terrestrial atmospheres and the rocky surface conditions of airless worlds. These topics introduce new opportunities for planetary scientists to engage with astronomers and expand science previously bound within the Solar System to a wider diversity of worlds. The community can also leverage the results from this program to identify those planets that warrant more detailed follow-up observations, stimulating GO proposals in future cycles. The findings from this program may also inform the community's strategy in developing future exoplanet programs with other facilities.

The $10^4$ hr Exoplanet Survey will help to more closely share expertise between the communities that study transiting exoplanets and those that study directly imaged substellar companions and isolated brown dwarfs. The full diversity of atmospheres cannot be fully understood without tying together the extreme ends of parameter space and the uniform sample of measurements envisioned in this survey will allow these two fields to more easily compare results and share knowledge. The Early Release Science programs in these two communities have already helped to share knowledge within each field, but the Survey could facilitate greater communication between the two. The cohesive dataset provided by the Survey should be an extraordinary resource for the community, which can be mined for science for years to come in addition to kickstarting many expanded investigations through GO proposals in future cycles.

We suggest that the implementation of these recommendations be informed by the opportunities for increased community engagement. The efforts should be built in ways that strategically work to optimize community collaboration and scientific output. In particular, we advocate that Archival GO programs may be one way to support fruitful connections between different disciplines and expertise, but recommend that STScI consider what other avenues exist and could be supported.

# 10. Conclusions

The rapid rate of discovery in exoplanets is likely to continue in the foreseeable future. *HST* has demonstrated its tremendous impact in exoplanet science and the initial cycles of *JWST* make it clear that it will be a profoundly impactful mission as well. New methods of exoplanet discovery and characterization are on the horizon. Formerly isolated communities of stellar astrophysics, exoplanets, and planetary science are beginning to overlap and connect, as are the direct imaging and transit observational communities. The next generation of missions and observing campaigns are being designed with the current knowledge of exoplanet populations. This is where the recommended DDT concept and exoplanet science support recommendations can profoundly shape the course of research in this field. A significant allocation of DDT time now on a comprehensive rocky planet survey can definitively answer a fundamental question that might otherwise take much of the *JWST* mission lifetime to address and kickstart community-driven followup observations on the most high value planets. Such a program, and others like it, as presented in the community WPs, will ultimately aggregate into an inevitable $10^4$ Hour Exoplanet Survey by the conclusion of the *JWST* mission. A small investment of structural support now to support the needs of long term, multi-epoch, multi-wavelength, high S/N observations will pay large dividends in a legacy exoplanet dataset that can be mined for generations of astronomers. Finally, the recommendations around funding support and proprietary policies will help to protect and encourage career development of early career researchers in our young field. It will ultimately be these individuals that will be leading the field at the conclusion of the *JWST* mission and transitioning the field to the next generation facility on the pathway to habitable worlds.

# Acknowledgements

The WG gratefully acknowledges all the engagement by the community and particularly the time and effort spent providing survey responses and writing white papers. These ideas were the basis of our recommendations and so these efforts could not have been successful without them. They also show the rich diversity of scientific questions yet to be answered in our field and demonstrate the unique capabilities that *JWST* and *HST* have to address them. The WG deeply appreciates all the support from STScI, from soliciting community input by constituting this WG, to logistical support for our meetings and work.

# Appendices

## A. White Papers

1. Standing on the Shoulders of Giants: A Comprehensive Spectroscopic Survey of Transiting & High-Contrast Giant Planets
   Alam, Munazza K.; Rickman, Emily; Hoch, Kielan; Mollière, Paul; Lothringer, Josh; Carter, Aarynn L.; Rebollido, Isabel; Sutlieff, Ben J.; Kammerer, Jens;

2. Statistical survey of intermediate mass planets
   Allart, Romain; Chachan, Yayaati; Cowan, Nicolas; Sikora, James; Radica, Michael; Lafreniere, David; Coulombe, Louis-Philippe; Doyon, René; Cook, Neil James; Loic Albert; Krishnamurthy, Vigneshwaran; Cadieux, Charles; Dang, Lisa,

3. Constraining the effect of stellar activity on exoplanet transmission spectra
   Allen, Natalie H.; Ahrer, Eva-Maria; Foote, Trevor;

4. Don’t Stop Believin’: The Importance of Continuing the Search for Atmospheres on Rocky Exoplanets with JWST
   Allen, Natalie; Bennett, Katherine; Dos Santos, Leonardo A.; Gressier, Amelie; Rustamkulov, Zafar;

5. Flexible Scheduling: A New Technique for Maximizing Scientific Output from Exoplanet Observations
   Bennett, Katherine; Dos Santos, Leonardo A.; Gressier, Amelie; Ramos-Rosado, Lakeisha; Rustamkulov, Zafar;

6. Contextualizing Planetary Systems and their Host Stars for the Habitable Worlds Observatory
   Bowler, Brendan; Teixeira, Katie; Franson, Kyle; Booth, Mark; Bowens-Rubin, Rachel; Carter, Aarynn; Cugno, Gabriele; Factor, Sam; Girard, Julien H.; Hoch, Kielan; Kammerer, Jens; Limbach, Mary Anne; Manjavacas, Elena; Martinez, Raquel A.; Patapis, Polychronis; Rebollido, Isabel; Rickman, Emily; Sutlieff, Ben J.; Theissen, Chris; Vos, Johanna; Zhang, Zhoujian; Zhou, Yifan;

7. Tilting Gas Giants: A New Window into Planet Formation
   Bryan, Marta L.; Bowler, Brendan P.; Zhou, Yifan; Millar-Blanchaer, Maxwell A.; Martinez, Raquel A.; Sutlieff, Ben J.; Adams, Arthur D.; Franson, Kyle;

8. Eta-Earth with the Hubble Space Telescope
   Burke, Christopher J.; Mullally, Susan E.; McCullough, P. R.; Christiansen, Jessie L.; Mulders, Gijs D.; Knicole D.; Dressing, Courtney; Agol, Eric; Eastman, Jason; Quintana, Elisa; Fernandes, Rachel B.; Ciardi, David R.; Kunimoto, Michelle; Huber, Daniel; Johnson, Samson; Weiss, Lauren; Kopparapu, Ravi; Zink, Jon; Kipping, David; Bergsten, Galen; Mandell, Avi M.; Hardegree-Ullman, Kevin K.;

9. The Unrealised Interdisciplinary Advantage of Observing High Mass Transiting Exoplanets and Brown Dwarfs
   Carter, Aarynn L.; Beatty, Thomas; Casewell, Sarah; Lewis, Nikole; Moran, Sarah E.; Wakeford, Hannah R.; Alam, Munazza; Chubb, Katy L.; Hoch, Kielan; Lothringer, Joshua D.; Manjavacas, Elena;

10. Investing in the Unrivaled Potential of Wide-Separation Sub-Jupiter Exoplanet Detection and Characterisation with JWST
    Carter, Aarynn L.; Bowens-Rubin, Rachel; Calissendorff, Per; Kammerer, Jens; Li, Yiting; Meyer, Michael R.; Booth, Mark; Factor, Samuel M.; Franson, Kyle; Gaidos, Eric.; Leisenring, Jarron M.; Lew, Ben W.P.; Martinez, Raquel A.; Rebollido, Isabel; Rickman, Emily; Sutlieff, Ben J.; Ward-Duong, Kimberly; Zhang, Zhoujian;

11. Cool Planets Legacy Survey
    Cassese, Ben; Kipping, David;

12. Temperate sub-Neptunes as a window on habitable worlds and the early Earth
    Charnay, Benjamin; Turbet, Martin; Kite, Edwin S.; Hu, Renyu;

13. A Panchromatic Survey of Warm and Hot Jupiters
    Cubillos, Patricio;

14. Sub-Neptune atmospheres in response to UV photons
    Diamond-Lowe, Hannah; Mendonça, João; Rathcke, Alexander;

15. Staring at habitable-zone terrestrial worlds
    Diamond-Lowe, Hannah; Rathcke, Alexander; Buchhave, Lars

16. In-Depth Atmospheric Characterization of the Temperate Water World Candidate LHS1140 b
    *Doyon, René; *Cadieux, Charles; Valencia, Diana; Plotnykov, Mykhaylo; Turbet, Martin; Fauchez, Thomas; Allart, Romain; Albert, Loïc; Artigau, Étienne; Canto, Bruno L.; Chachan, Yayaati; Cherubim, Collin; Cloutier, Ryan; Cook, Neil James; Cowan, Nicolas; Dang, Lisa; Delfosse, Xavier; Dumusque, Xavier; Ehrenreich, David; Jahandar, Farbod; Jayawardhana, Ray; Kaltenegger, Lisa; Krishnamurthy, Vigneshwaran; Lafrenière, David; Leão, Izan; L'Heureux, Alexandrine; Lim, Olivia; Lima, Roseane; MacDonald, Ryan; Messias, Yuri S.; Piaulet, Caroline; Rackham, Benjamin V.; Radica, Michael; Renan de Medeiros, José; Rowe, Jason; Salhi, Salma; Taylor, Jake;

17. Witnessing the Evolution of Sub-Neptunes
    Feinstein, Adina; Thao, Pa Chia; Mann, Andrew; Welbanks, Luis; Barat, Saugata; Fernandes, Rachel; Gao, Peter; Levine, W. Garrett; Luque, Rafael; Nine, Andrew; Rockcliffe, Keighley; Schulte, Jack; Seligman, Darryl; Sikora, James; Soares-Furtado, Melinda; Vissipragada, Shreyas;

18. A Dynamically Informed Planet Search: Using Astrometric Accelerations to Directly Image Mature Giant Planets
    Franson, Kyle; Bowler, Brendan P.; Teixeira, Katie; Balmer, William O.; Carter, Aarynn; Factor, Samuel M.; Rickman, Emily; Bowens-Rubin, Rachel; Girard, Julien H.; Cugno, Gabriele; Marino, Sebastian; Kammerer, Jens; Vos, Johanna; Sutlieff, Ben J.; Theissen, Christopher A.; Zhang, Zhoujian;

19. Money, Money, Money
    Fu, Guangwei; Mansfield, Megan; Kempton, Eliza; Bean, Jacob; Fortney, Jonathan; Schlawin, Everett;

20. Maximizing the Exoplanet Potential of the Roman Galactic Bulge Time Domain Survey via HST and JWST Precursor Imaging
    Gaudi, B. Scott; Bennett, David P.; Mroz, Przemek; Nataf, David M.; Penny, Matthew; Quintana, Elisa V.; Wilson, Robert F.; Terry, Sean K.,

21. Constraining the habitability of rocky exoplanets in the Milky Way
    Greene, Thomas;

22. A Best-in-Class Statistical Survey of Exoplanet Atmospheres
    Eliza Kempton; Jegug Ih; Jacob Bean; Drake Deming; Jonathan Fortney; Mike Line; Tom Mikal-Evans; Jasmine Blecic; David Ciardi; Lia Corrales; Nicolas Cowan; Tansu Daylan; René Doyon; Billy Edwards; Guangwei Fu; Peter Gao; Renyu Hu; Lisa Kaltenegger; Thaddeus Komacek; David Latham; Megan Mansfield; Mercedes Lopez-Morales; Rafael Luque; Caroline Morley; Norio Narita; Matthew Nixon; Everett Schlawin; Alessandro Sozetti; Keivan Stassun; Jeff Valenti; Ian Wong

23. Exomoons: Their Scientific Potential, Urgent Need and JWST's Unique Opportunity
    Kipping, David M.; Martinez, Miguel A. S.; Sucerquia, Mario.; Quarles, Billy L.; Li, Gongjie; Teachey, Alex; Tinetti, Giovanna; Szulagyi, Judit; Yahalomi, Daniel A.; Trani, Alessandro A.; Saha, Suman; Raymond, Sean N.; Sutton, Phil J; Rein, Hanno; Hippke, Michael; Satyal, Suman; Canup, Robin; Rosario-Franco, Marialis; Sandford, Emily; Cassese, Ben; Dalba, Paul A.; Fabrycky, Daniel C.; Perets, Hagai B; Sengupta, Sujan; Gordon, Tyler A.; Saillenfest, Melaine; Zuluaga, Jorge I.; Moraes, Ricardo A.; Haqq-Misra, Jacob; Schmidt, Carl A.; Narang, Mayank; Veras, Dimitri; Trierweiler, Isabella; Holman, Matthew J.; Tokadjian, Armen; Baillié, Kevin; Alvarado-Montes, J. A.; Szabó M, Gyula; Rodriguez, Joseph E.; Sulis, Sophia; Haghighipour, Nader; Turner, Edwin L.; Kenworthy, Matthew A.; Roccetti, Giulia; Nesvorny, David; Lillo-Box, Jorge; Donnison, John Richard; Kleisioti, Evangelia; Hill, Michelle L.; Agol, Eric; Efroimsky, Michael;

24. A Search for Life Around Dead Stars
    Limbach, Mary Anne; Vanderburg, Andrew; Adams, Elisabeth; Agol, Eric; Becker, Juliette; Blouin, Simon; Chen, Howard; Clement, Matt; Cunningham, Tim; Debes, John; Fauchez, Thomas; Haqq Misra, Jacob; Heller, René; Janson, Markus; Kaltenegger, Lisa; Kenworthy, Matthew; Kilic, Mukremin; Kleisioti, Evangelia; Kopparapu, Ravi; Livesey, Joseph; López-Morales, Mercedes; MacDonald, Ryan; Mahadevan, Suvrath; Mullally, Susan; Quintana, Elisa; Schwieterman, Edward; Shields, Aomawa; Soares-Furtado, Melinda; Steckloff, Jordan; Stefansson, Gudmundur; Stevenson, Kevin; Veras, Dimitri; Yarza, Ricardo;

25. Atmospheric Erosion along the Cosmic Shoreline
    Lustig-Yaeger, Jacob; Stevenson, Kevin B.; Bennett, Katherine; Mansfield, Megan; King, George W.; Ih, Jegug; Allen, Natalie H.;

26. The Search for Life on Exoplanets with JWST
    Madhusudhan, Nikku;

27. Mind the Gap: Crunching the Atmospheres of the Coldest Free-Floating Brown Dwarfs
    Manjavacas, Elena; Metchev, Stan; Girard, Julien H; Theissen, Cris; Kammerer, Jens; Sutlieff, Ben J.; Hinkley, Sasha; Bowler, Brendan; Vos, Johanna; Martinez, Raquel A.; Suárez, Genaro; Zhou Yifan; Hoch, Kielan; Factor, Samuel M.; Zhoujian Zhang; Rickman, Emily; Franson, Kyle; Leggett, Sandy K.; Morley, Caroline V.; Karalidi, T.;

28. Preliminary Detection of a Terrestrial Exoplanet Atmosphere with JWST
    Mansfield, Megan; Zhang, Michael; Bean, Jacob L.; Xue, Qiao; Stevenson, Kevin; Lustig-Yaeger, Jacob; Kite, Edwin S.; Fu, Guangwei; Ih, Jegug;

29. The Astrobiology of The TRAPPIST-1 System
    Meadows, Victoria; Lincowski, Andrew; Lustig-Yaeger, Jacob; Krissansen-Totton, Joshua; de Wit, J.; Doyon, René ; Gillon, Michael;

30. No title listed
    May, Erin; Hammond, Mark; Stevenson, Kevin; Komacek, Thaddeus; Lothringer, Joshua; Schlawin, Everett; Mayorga, Laura; Alam, Munazza; Gao, Peter; Moran, Sarah; Wakeford, Hannah; Lee, Elspeth; Parmentier, Vivien;

31. Time-monitoring of “Extreme” Debris Disks
    Moro-Martin, Amaya; Chen, Christine; Su, Kate Y. L.; Worthen, Kadin;

32. Constraining Atmospheric Mixing in Gas Giants with JWST
    Mukherjee, Sagnick; Fortney, Jonathan J.; Ohno, Kazumasa;

33. Searching for Jovian Analogs Around White Dwarf Stars
    Mullally, Susan E.; Debes, John; Kuchner, Marc; Fleming, Scott; Girard, Julian; Mulders, Gjis D.; Kammerer, Jens; Kilic Mukremin; Grunblatt, Samuel K.; Quintana, Elisa V.; MacDonald, Ryan J.; Albert, Loic,;Hermes, JJ; Reach, William T.; Colon, Knicole; Hines, Dean C.; von Hippel, Ted; Soares-Furtado, Melinda; Montet, Benjamin; Mullally, Fergal; Burke, Christopher J.; Rebollido, Isabel Rebollido; Rickman, Emily; Factor, Samuel M.; Dos Santos, Leonardo

34. Leveraging the Pandora SmallSat Mission for Exoplanet Science with HST and JWST
    Rackham, Benjamin V.; Quintana, Elisa V.; Colón, Knicole D.; Allen, Natalie H.; Ciardi, David R.; Foote, Trevor; Fuda, Nguyen; Greene, Thomas; Hedges, Christina; Hoffman, Kelsey L.; Lewis, Nikole K.; Mullally, Susan E.; Mann, Andrew W.; Mansfield, Megan; Mason, James Paul; Morris, Brett M.; Rowe, Jason, F. ;

35. No title listed
    Radica, Michael; Alderson, Lili;

36. The Search for Rocky Exoplanet Atmospheres
    Stevenson, Kevin B.; Lustig-Yaeger, Jacob; Mansfield, Megan; Bennett, Katherine; King, George W.; Ih, Jegug; Allen, Natalie H.;

37. Prioritizing High-Precision Photometric Monitoring of Exoplanet and Brown Dwarf Companions with JWST
    Sutlieff, Ben J.; Chen, Xueqing; Liu, Pengyu; Bubb, Emma E.; Metchev, Stanimir; Bowler, Brendan; Vos, Johanna; Martinez, Raquel A.; Suárez, Genaro; Zhou, Yifan; Factor, Samuel M.; Zhang, Zhoujian; Rickman, Emily; Adams, Arthur D.; Manjavacas, Elena; Girard, Julien H.; Kim, Bokyoung; Dupuy, Trent J.;

38. Next step on the roadmap to the efficient and robust characterization of temperate terrestrial planet atmospheres with JWST
    TRAPPIST-1 JWST Community Initiative; *de Wit, J.; *Doyon, Renè ; Rackham, Benjamin V.; Lim, Olivia; Ducrot, Elsa; Kreidberg, Laura; Benneke, Björn; Ribas, Ignasi; Berardo, David; Niraula, Prajwal; Iyer, Aishwarya; Shapiro, Alexander; Kostogryz, Nadiia; Witzke, Veronika; Gillon, Michael; Agol, Eric; Meadows, Victoria; Burgasser, Adam J.; Owen, James E.; Fortney, Jonathan J.; Selsis, Franck; Bello-Arufe, Aaron; Bolmont, Emeline; Cowan, Nicolas; Dong, Chuanfei; Drake, Jeremy J.; Garcia, Lionel; Greene, Thomas; Haworth, Thomas; Hu, Renyu; Kane, Stephen R.; Kervella, Pierre; Koll, Daniel; Krissansen-Totton, Joshua; Lagage, Pierre-Olivier; Lichtenberg, Tim; Lustig-Yaeger, Jacob; Lingam, Manasvi; Turbet, Martin; Seager, Sara; Barkaoui, Khalid; Bell, Taylor J.; Burdanov, Artem; Cadieux, Charles; Charnay, Benjamin; Cloutier, Ryan; Cook, Neil J.; Correia, Alexandre C. M.; Dang, Lisa; Daylan, Tansu; Delrez, Laetitia; Demory,

Brice-Olivier; Edwards, Billy; Fauchez, Thomas J.; Flagg, Laura; Fraschetti, Federico ; Haqq-Misra, Jacob; Huang, Ziyu; Iro, Nicolas; Jayawardhana Ray; Jehin, Emmanuel; Jin, Meng; Kite, Edwin; Kitzmann, Daniel; Kral, Quentin; Lafrenière, David; Libert, Anne-Sophie; Liu, Beibei; Mohanty, Subhanjoy; Morris, Brett M.; Murray, Catriona A.; Piaulet, Caroline; Pozuelos, Francisco J.; Radica, Michael; Ranjan, Sukrit; Rathcke, Alexander; Roy, Pierre-Alexis; Schwieterman, Edward W.; Turner, Jake D.; Triaud, Amaury; Ward, Howard S.; Way, Michael J.;

39. Know Thy Star, Know Thy Planet: In-Depth Stellar Activity Characterization of Trappist-1
    TRAPPIST-1 JWST Community Initiative; *Doyon, René ; *de Wit, J.; Rackham, Benjamin V.; Lim, Olivia; Ducrot, Elsa; Ribas, Ignasi; Berardo, David; Niraula, Prajwal; Iyer, Aishwarya; Shapiro, Alexander; Kostogryz, Nadiia; Witzke, Veronika; Gillon, Michael; Agol, Eric; Meadows, Victoria; Burgasser, Adam J.; Owen, James E.; Fortney, Jonathan J.; Selsis, Franck; Bello-Arufe, Aaron; Bolmont, Emeline; Cowan, Nicolas; Dong, Chuanfei; Drake, Jeremy J.; Garcia, Lionel; Greene, Thomas; Haworth, Thomas; Hu, Renyu; Kane, Stephen R.; Kervella, Pierre; Koll, Daniel; Krissansen-Totton, Joshua; Lagage, Pierre-Olivier; Lichtenberg, Tim; Lustig-Yaeger, Jacob; Lingam, Manasvi; Turbet, Martin; Seager, Sara; Barkaoui, Khalid; Bell, Taylor J.; Burdanov, Artem; Cadieux, Charles; Charnay, Benjamin; Cloutier, Ryan; Cook, Neil J.; Correia, Alexandre C. M.; Dang, Lisa; Daylan, Tansu; Delrez, Laetitia; Demory, Brice-Olivier; Edwards, Billy; Fauchez, Thomas J.; Flagg, Laura; Fraschetti, Federico ; Haqq-Misra, Jacob; Huang, Ziyu; Iro, Nicolas; Jayawardhana Ray; Jehin, Emmanuel; Jin, Meng; Kite, Edwin; Kitzmann, Daniel; Kral, Quentin; Lafrenière, David; Libert, Anne-Sophie; Liu, Beibei; Mohanty, Subhanjoy; Morris, Brett M.; Murray, Catriona A.; Piaulet, Caroline; Pozuelos, Francisco J.; Radica, Michael; Ranjan, Sukrit; Rathcke, Alexander; Roy, Pierre-Alexis; Schwieterman, Edward W.; Turner, Jake D.; Triaud, Amaury; Ward S. Howard; Way, Michael J.;

40. Essential X-rays and UV characterization of stars that host exoplanets of high interest
    Aline A. Vidotto; Allison Youngblood; Amélie Gressier; Antonija Oklopčić; Antonio García Muñoz; David Ehrenreich; Evgenya L. Shkolnik; George King; Girish M. Duvvuri; James Kirk; James E. Owen; Katherine Bennett; Lakeisha Ramos-Rosado; Leonardo A. Dos Santos; Mercedes López-Morales; Munazza Alam; Natalie Allen; Patrick McCreery; R. O. Parke Loyd; Shreyas Vissapragada; Vincent Bourrier;

41. Variability in Disks; a Signpost for Exoplanets
    Wolff, Schuyler G.; John Debes; Millar-Blanchaer, Maxwell A.; Leisenring, Jarron; Rickman, Emily; Rebollido, Isabel; Metchev, Stanimir; Gaspar, Andras;

42. Calibrating Accretion Indicators in the Planetary-mass Regime with Multi-epoch UV and Hα Observations
    Zhou, Yifan; Bowler, Brendan; Balmer, William; Girard, Julien; Cugno, Gabriele; Martinez, Raquel A.; Franson, Kyle; Bryan, Marta;

# B. Survey

## Section 1 - Demographic Information

1. What type of institution are you primarily affiliated with?
    - Doctoral degree granting university
    - Other four-year university or college
    - Two-year college or community college
    - Observatory, Laboratory, or National Facility
    - Research institution
    - Other:

2. Please check all that apply.
    - My primary affiliation is an R1 or equivalent institution
    - My primary affiliation is primarily an undergraduate institution
    - My primary affiliation is a Minority Serving Institution in the US (MSI; https://msiexchange.nasa.gov/).
    - My primary affiliation is a government or research institution,
    - Other:

3. What geographic region is your primary institution located in?
    - Africa
    - Asia
    - Australasia
    - Central or South America
    - Europe
    - North America, not in the United States
    - United States

4. What sub-field(s) do you do most of your work in?

5. What is your career stage?
    - Undergraduate student
    - Graduate student (PhD or Masters)
    - Postdoc
    - Research scientist or similar long-term position
    - Tenure-track or early career permanent position
    - Tenured or established permanent position
    - Other:

6. When did you get your PhD? (Note that "not applicable" and "not yet" are both valid options.)

7. What is your gender identity?
   - Female
   - Male
   - Non-binary
   - Prefer not to say
   - Other:

8. What do you identify as your race and/or ethnicity?
   - American Indian or Alaskan Native
   - Asian
   - Black or African American
   - Hispanic or Latinx
   - Native Hawaiian or Pacific Islander
   - White
   - Prefer not to respond
   - Other:

9. What involvement have you had in proposing for the use of HST or JWST (regardless of the outcome)
   - PI of HST or JWST proposal
   - Major contributing Co-I to HST or JWST proposal
   - Minor contributing Co-I to HST or JWST proposal

10. Have you been involved in the analysis of new or archival HST or JWST data?
    - Led the analysis or interpretation of data
    - Collaborated on the analysis or interpretation of data

## Section 2 - Using HST and JWST to study planetary systems:

1. Key Science Themes

   Please respond to one or more of the following questions (1750 character limit):

   - How can HST and JWST be used to maximize exoplanet science?
   - What science topics or questions should be prioritized for observations?
   - What science areas need to be prioritized for HST and JWST to lay the foundation for future flagships?
   - What science could best be enabled by joint, coordinated observations from both HST and JWST?

2. Optimal Timing

   Please respond to one or more of the following questions (1750 character limit):

   - What timing considerations affect the highest priority science areas for HST and JWST (e.g., simultaneous, reconnaissance/pilot observations)?
   - How can these timing considerations best be implemented?

3. Scale of Resources:

   Please respond to one or more of the following questions (1750 character limit):

   - What is the appropriate scale of resources required to support exoplanet science with HST and/or JWST (e.g., number of orbits/hours, financial support, scheduling priority, etc.)?
   - What resources are necessary to facilitate coordinated observations from HST and JWST?
   - What other facilities would be needed to support this work (e.g., telescope, computing, laboratory)?

4. Director Discretionary Time:

   Please respond to one or more of the following questions (1750 character limit - Detailed discussions should be made in the form of a white paper):

   - What is the appropriate scale of resources required to support exoplanet science with HST and/or JWST (e.g., number of orbits/hours, financial support, scheduling priority, etc.)?
   - What resources are necessary to facilitate coordinated observations from HST and JWST?
   - What other facilities would be needed to support this work (e.g., telescope, computing, laboratory)?

## Section 3 - Additional feedback:

You are invited to use this space for any additional thoughts regarding optimal strategies for maximizing the scientific return from HST and JWST spectroscopic and imaging observations of planetary systems.

1. Is there anything we have missed in the above questions?

   (1750 character limit)